\documentclass[12pt]{article}
\usepackage[top=1in, bottom=1in, left=1.in, right=1.in]{geometry}
\usepackage[english]{babel}
\usepackage{atbegshi,cite}
\usepackage{amsmath,amssymb,amsbsy,amstext, amsthm, simplewick}
\usepackage{braket}
\usepackage{hyperref}
\usepackage{graphicx}
\usepackage{amsfonts}
\usepackage[small]{caption}
\usepackage{upgreek}
\usepackage[titletoc]{appendix}
\usepackage[usenames,dvipsnames,table]{xcolor}
\usepackage{setspace}
\usepackage{color}
\usepackage{quantikz}
\usepackage{comment}

\newcommand{\newc}{\newcommand}
\newc{\TR}{{\rm Tr}}
\newc{\calA}{{\cal A}}
\newc{\calH}{{\cal H}}
\newc{\calO}{{\cal O}}

\begin{document}

\begin{titlepage}
\begin{flushright}
{\large 
~\\
}
\end{flushright}

\vskip 2.2cm

\begin{center}

{\large \bf Block Encoding Non-Abelian Lattice Gauge Theory}

\vskip 1.4cm
Patrick Draper
\\
\vskip 1cm
{ Department of Physics, University of Illinois, Urbana, IL 61801}
\vspace{0.3cm}
\vskip 4pt

\vskip 1.5cm

\begin{abstract}
Gauge theories like lattice QCD present a complex problem for quantum simulation. In a basis where the electric part of the Hamiltonian is simple, the magnetic part, generally expressed as a sum over the plaquette operators of the lattice, is quite complicated, producing correlated transitions between several link and site degrees of freedom. We provide an efficient block encoding of the plaquette operator in the irrep basis, a refinement of the electric basis where the internal gauge-variant degrees of freedom are integrated out. The construction removes the plaquette matrix element scaling wall which has been a significant barrier for other approaches in this basis. The algorithm leverages a convenient factorization property of the matrix elements, cheap classical precomputation, and quantum oracles built from lookup tables and programmed rotations. 
\end{abstract}

\end{center}

\vskip 1.0 cm

\end{titlepage}

\setcounter{footnote}{0} 
\setcounter{page}{1}
\setcounter{section}{0} \setcounter{subsection}{0}
\setcounter{subsubsection}{0}
\setcounter{figure}{0}

\section{Introduction}
\label{sec:intro}

Lattice gauge theory is an important scientific target for quantum simulation.  Gauge theories form the backbone of the Standard Model of particle physics, describing nature's fundamental quantum mechanical degrees of freedom and interactions down to the shortest accessible distance scales.  For decades, classical Monte Carlo simulations of  lattice quantum chromodynamics (QCD) in equilibrium have provided deep insight into the fundamental physics of the strong interactions.  Achieving large-scale quantum simulations of lattice gauge
theories will open access to the  new regime of real-time gauge field dynamics, where strong quantum correlations must be tracked through a nonequilibrium system of many degrees of freedom. Historically, the lattice QCD program played an important role in setting ambitious goals and driving forward the field of high-performance computing, and quantum simulations of QCD are now positioned to play the same role for quantum computing.

At present, we are in what might optimistically be termed the ``late NISQ era.''  Quantum simulation of gauge theories has matured rapidly over the past
decade, developing from the first real-time simulations of $1+1$D quantum electrodynamics~\cite{martinez2016} to increasingly ambitious algorithmic
developments and hardware implementations for non-Abelian theories.  An incomplete survey of milestones includes the first digital implementations of $\mathrm{SU}(2)$ in one spatial dimension on superconducting hardware~\cite{klco2020su2}; the development of  different encoding schemes, including the  local multiplet/irrep/reduced electric basis~\cite{banuls2017su2basis,ciavarella2021trailhead,balaji2025circuits}, the loop-string-hadron basis~\cite{raychowdhury2020gausslaw,raychowdhury2020lsh,kadam2023lsh_su3,kadam2025lsh_trivalent}, the group element basis~\cite{lamm2019,gustafson2024su3subgroup}, and the orbifold approach~\cite{hanada2021orbifold,bergner2024orbifoldqcd,bergner2025orbifoldks}; and the constructino of  resource-efficient algorithms for simulation~\cite{davoudi2023algs,muller2023yao,lamm2024block,gustafson2024su3subgroup,balaji2025circuits,balaji2025trunc,ymcirc}. 
 Demonstrations of  dynamics range from pair creation in the Schwinger model on trapped ions~\cite{nguyen2022schwinger} to thermalization in cold-atom gauge simulators~\cite{zhou2022thermalization,halimeh2023coldatom}, a program surveyed together with the broader out-of-equilibrium frontier in~\cite{halimeh2025outofequil}. Studies of larger systems and higher dimensions include $100{+}$-qubit Schwinger-model wavepacket dynamics~\cite{farrell2024schwinger}, hardware execution of $d=2$ $\mathrm{SU}(3)$ circuits using the large $N$ expansion~\cite{ciavarella2024largeN}, and development and noiseless simulation of quantum circuits for $3+1$D QCD~\cite{hidalgo2026lqcd}.    Community summaries and roadmaps appear in~\cite{bauer2023prxq,bauer2023nrp,funcke2023review}.  Together, these works define a late-NISQ era of sufficient scale and fidelity to validate formulations, exercise error-mitigation pipelines, and probe real-time dynamics inaccessible to classical Monte Carlo.  

However, real-time dynamics at physically relevant scales (continuum-limit lattice spacings, large $d=3$ volumes, energy densities approaching $\Lambda_{\rm QCD}$,  evolution times long enough to resolve hadronization or scattering) demand circuit depths far beyond what noisy hardware can reliably execute, even with aggressive error mitigation.  Controlling errors at the accuracy needed for quantitative predictions ultimately requires fault-tolerant, error-corrected quantum computation. The present work is an exercise in algorithm design and resource estimation for that regime.\footnote{Noisy classical emulations and hardware demonstrations of simple physics simulations with a degree of error detection or correction have begun to appear, including, for example, treatments of the lattice Schwinger model~\cite{froland2026utility} in iceberg  codes~\cite{self2024protecting} and the $XY$ model  with $\calO(10^2)$ error-detected logical qubits~\cite{dasu2026computing}.}

The main new contribution we make is an explicit construction of the block-encoding oracle for the plaquette operator appearing as the magnetic density in non-Abelian lattice gauge theory Hamiltonians, in the sparse encoding framework of~\cite{berry2015sparse,low2019qubitization,gilyen2019qsvt}. The circuits are based heavily on quantum lookups of gauge theory data which is efficient to classically pre-compute using public tools.  We work in the irrep basis~\cite{banuls2017su2basis,ciavarella2021trailhead,balaji2025circuits}, which is gauge invariant and qubit-efficient, and our approach avoids the need for in-register Clebsch-Gordan arithmetic. A direct extension of previous work~\cite{balaji2025circuits,balaji2025trunc} would classically pre-compute the matrix elements of the plaquette operator and implement transitions one at a time via Givens rotations~\cite{ciavarella2021trailhead}. This approach is difficult to scale beyond the lowest truncations due to an explosion in the number of matrix elements. Instead, our construction exploits the factorization of each plaquette matrix element into four ``site factors"~\cite{balaji2025circuits} which reduces the number of distinct lookup-table entries by many orders of magnitude.  The algorithm we describe is quite general, but specific numbers are given for gauge group SU(3), the gluon interactions of QCD. The formalism we develop can be extended straightforwardly to full QCD, where the gauge-matter interaction terms also decompose into site factors in the irrep basis~\cite{hidalgo2026lqcd}. 

Fault-tolerant resource estimation for lattice gauge theories has itself become a benchmark problem.  Early Trotter-based analyses by Kan and Nam~\cite{kan2021lqcd} gave explicit fault tolerant $T$-gate counts for gauge theories in general spatial dimension, but their $\mathrm{SU}(3)$ $T$-gate count estimates are prohibitively enormous.  Rhodes, Kreshchuk, and Pathak~\cite{rhodes2024exponential} subsequently replaced Trotterization with sparse block-encoding oracles and qubitization, reducing the computational cost by many orders of magnitude. They outline a $\mathrm{SU}(3)$ magnetic block encoding based on Clebsch-Gordan arithmetic but do not give an explicit construction, and their cost estimates are  based on  analytic cutoff-scaling formulas rather than precomputation.  We provide a complete construction that avoids in-register CG computation, and we obtain concrete per-query
$T$-counts that capture the actual sparsity and  structure of the Hamiltonian at physically relevant truncations.  On the matter side, Ref.~\cite{lamm2026chiral} recently gave a quantum signal processing implementation of the gauge-matter sector with exact chiral symmetry, taking the gauge-field oracles as given; the present work supplies the complementary magnetic piece. 

A condensed summary of the resource cost we obtain for SU(3) lattice gauge theory is as follows. We consider two truncations, both including up to the two-index tensors $\{1,3/\bar 3, 8, 6/\bar 6\}$ in the irrep catalog, and cutting off the per-vertex total electric Casimir at  $\leq 6$ and $\leq 9$, respectively.  A single call to the plaquette $\Box_p+\Box_p^\dagger$ oracle at these truncations costs about $10^5$ or $10^6$ $T$ gates. We do not attempt to fully optimize these numbers and it is likely that they can be substantially lowered. However, for a qualitative comparison, a single Trotter step on a plaquette implemented by the Givens rotation approach of~\cite{ciavarella2021trailhead,balaji2025circuits,balaji2025trunc} would, based on raw $\Box_p$ matrix element counts at these truncations, incur a cost at least a factor of $10^5$ higher. (It is also  difficult to classically pre-compute and store the matrix elements at this scale, whereas the pre-compute even for much milder truncations is easy in our approach.) 

Extending to a full periodic $L^3$ lattice, time evolution can be implemented with QSVT or a hybrid Trotter-QSVT algorithm. In the QSVT case, for lattices larger than $L\approx 20$, the dominant cost is the polynomial degree $\times$ routing, which is only mildly related to the plaquette oracle implementation through the subnormalization. With bare lattice coupling $g=1$,  the magnetic block encoding subnormalization is $3L^3\alpha_p$ with $\alpha_p\approx56$ per plaquette.\footnote{This number can be roughly estimated from the $\calO(1)$ matrix elements.  Our oracle is constructed in a way that the minimum subnormalization is the maximum row-1-norm of the operator. States couple directly to  tens of others at this truncation, so $\alpha_p$ lands at about fifty.} Routing costs another power of $L^3$. The unfavorable total volume-squared scaling remains a significant open problem. It suggests that detailed investigation of the hybrid algorithm would be interesting, but we do not carry it out in this work. 

The logical qubit count  is estimated as $Q\approx 12L^3+\calO(10^3)$. It is dominated by the register holding the lattice state. The additive term is working space (lookup workspace, flags, routing scratch, etc.) that does not scale with the volume, up to address registers of width $\log L$.

This work is organized as follows. In \S\ref{sec:lattice} we review the Kogut-Susskind Hamiltonian and the irrep / reduced-electric basis used as the computational basis. In   \S\ref{sec:blockencoding} we describe the decomposition of the full block encoding into electric and magnetic parts, and  the magnetic block encoding is described at a high level in \S\ref{sec:spo}. It is built primarily from a weighted prepare oracle (\S\ref{sec:oracle}), which produces a superposition over the neighbor states to which the plaquette operator transitions, with  amplitude controlled by the magnitude of the matrix element.  Full assembly of the magnetic block encoding is completed in \S\ref{sec:finishing}.  The main circuit elements are programmed rotation trees fed by quantum lookups, which read each plaquette state and return the classically pre-computed transition data in the form of rotation angles.  These primitives are reviewed in \S\ref{sec:data-quantum}.  Estimated $T$ gate costs for one call to the oracle  and logical qubit costs for the lattice state plus ancillas are computed in \S\ref{sec:cost}. In \S\ref{sec:qsvt-vs-trotter} we discuss usage of the oracle in QSVT and hybrid QSVT-Trotter simulations.

\S\ref{sec:conclusions} concludes with some directions for future work. A primary conclusion of this work is that our oracle efficiently mitigates the cost that scales with the count of nonzero plaquette matrix elements, which grows extremely rapidly with truncation.  Due to the residual volumetric scalings, the most physically demanding regimes remain an important target for future algorithmic development.

\section{Encoding Non-abelian Lattice Gauge Theory}
\label{sec:lattice}

The Kogut-Susskind Hamiltonian for pure Yang-Mills gauge theory on a spatial lattice is given by~\cite{kogut1975hamiltonian},
\begin{equation}
H = \sum_{\vec s}\left[ \frac{g^2}{2 a} \sum_{i} E^2(\vec s,\vec e_i)
+ \frac{1}{g^2 a} \sum_{i<j}\bigl(2N - \Box_p(\vec s,\vec e_i,\vec e_j) - \Box_p^\dagger(\vec s,\vec e_i,\vec e_j)\bigr)\right],
\label{eq:KS-Hamiltonian}
\end{equation}
where the magnetic term is built from the plaquette operator,
\begin{equation}
\Box_p = \TR\bigl[U_{\ell_1} U_{\ell_2} U_{\ell_3}^\dagger U_{\ell_4}^\dagger\bigr].
\label{eq:plaq-unitary}
\end{equation}
 $U_\ell$ is the link operator conjugate on link $\ell$. The left and right ends of the link host electric field operators $E_{L,R}$ conjugate to $U$, with commutation relations 
\begin{equation}
[E^a_L,U] = U\,T^a,\qquad
[E^a_R,U] = T^a U,\qquad
[E^a_{L},E^b_{L}] = i f^{abc} E^c_{L},\qquad
[E^a_{R},E^b_{R}] = -i f^{abc} E^c_{R},
\label{eq:link-algebra}
\end{equation}
where $T^a$ are the group generators in the fundamental representation, $f^{abc}$ are the structure constants, $[E^a_L,E^b_R]=0$, and the quadratic Casimir is $E^2 = E^a_L E^a_L = E^a_R E^a_R$.  Operators on different links commute.
We set the lattice spacing $a=1$ below, and we take an isotropic lattice of side length $L$ with periodic boundary conditions, for which $N_{\rm links}=N_{\rm plaq}=3L^3$.

The computational basis we use in this work is a variant of the electric basis, in which $E^2$ is diagonal and the magnetic term generates transitions. The first comprehensive treatment of the electric basis for quantum simulations was given by Byrnes and Yamamoto~\cite{byrnes2006simulating}. More recent work introduced a variant called the ``irrep" or ``local multiplet" basis, in which the internal color indices are eliminated~\cite{banuls2017su2basis,klco2020su2,ciavarella2021trailhead}. The simplification is possible because every physical state is gauge invariant, satisfying a Gauss law at each vertex. The physical electric basis states can be labeled purely by a choice of irreducible representation on each link and a  singlet multiplicity index at each vertex~\cite{balaji2025circuits}.
The latter is present to disambiguate cases where there is more than one distinct gauge invariant wavefunction in the tensor product of link irreps meeting at the vertex. It answers the question, ``given these link irreps, into which singlet are they fused?" given some agreed up canonical ordering. For low truncations and special lattices (only the fundamental and antifundamental irreps, with three links meeting at each vertex) the singlet is always unique, and the singlet multiplicity index can be dropped.  For higher truncations ($\geq$ two-index tensors with any type of vertex) or higher dimensions ($\geq$ one-index tensors with $d\geq 2$) the multiplicity index is an extra local degree of freedom that must be included. For example, in SU(3) gauge theory $8\otimes 8\otimes 8$ and $3\otimes \bar 3 \otimes 3 \otimes \bar 3$ both contain two distinct singlets in their direct sum decomposition. The combination of the link irrep and site singlet data was termed the ``reduced electric basis" in~\cite{balaji2025circuits}, and is the basis used in this work. 

The number of irreps is infinite, so to encode the theory on a quantum computer we must truncate the Hilbert space. A simple choice, particularly effective for studying low energy physics in the strong coupling limit, is to keep only link irreps with quadratic Casimir below a cutoff, a per-link electric energy cut. A variant is to cut on the  electric energy associated with a site, defined as the sum of all link Casimirs meeting at a lattice site~\cite{balaji2025trunc},
\begin{equation}
\sum_{\ell(s)}C_2(\ell(s))\leq B,
\end{equation}
for some cutoff $B$. Further variants with more sculpted truncations on the states may be advantageous, or approaches based on other controlled approximations~\cite{ciavarella2024largeN}. In this work we will use $B=6$ and $B=9$ in SU(3) gauge theory as examples. Both truncations include the link irrep set $\{1,3/\bar 3, 8, 6/\bar 6\}$, or the one and two-index tensors, which carry Casimirs $\{0,4/3,3,10/3\}$ respectively. They differ in what link excitations can meet at a vertex. With $B=6$, for example, four triplets, or, two adjoints, or two triplets and a sextet can meet at a vertex. With $B=9$ up to three adjoints can meet at a vertex, and so on. Higher-rank irreps are excluded by the Casimir cutoff at $B=6$ and $B=9$, but appear at $B=10$. The physics reach of the minimal ($1,3,\bar 3$) truncation has recently been examined in the context of the string tension~\cite{chen2026minimal}. What is actually needed for a faithful simulation of continuum gauge theory is a subject of ongoing work~\cite{ciavarella2025truncations,drapergupta2026}; we anticipate a qualitative level of accuracy at the truncations considered here, and our oracle construction below can accommodate higher truncations required for quantitative precision.

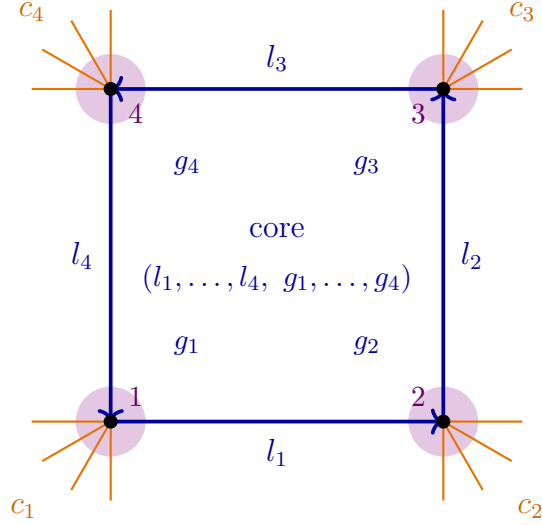
\begin{figure}[t]
\centering
\begin{tikzpicture}[scale=1.1]
  \foreach \p in {(0,0),(4,0),(4,4),(0,4)} \fill[violet!22] \p circle (0.42);
  \draw[line width=1.4pt,blue!60!black,->] (0,0) -- node[below=2pt]{$l_1$} (4,0);
  \draw[line width=1.4pt,blue!60!black,->] (4,0) -- node[right=2pt]{$l_2$} (4,4);
  \draw[line width=1.4pt,blue!60!black,->] (4,4) -- node[above=2pt]{$l_3$} (0,4);
  \draw[line width=1.4pt,blue!60!black,->] (0,4) -- node[left=2pt]{$l_4$} (0,0);
  \def\R{0.95}
  \foreach \a in {90,120,150,180}  \draw[orange!90!black,thick] (0,4) -- ++(\a:\R);
  \foreach \a in {180,210,240,270} \draw[orange!90!black,thick] (0,0) -- ++(\a:\R);
  \foreach \a in {270,300,330,360} \draw[orange!90!black,thick] (4,0) -- ++(\a:\R);
  \foreach \a in {0,30,60,90}      \draw[orange!90!black,thick] (4,4) -- ++(\a:\R);
  \node[orange!80!black] at (-1.05,-1.05) {$c_1$};
  \node[orange!80!black] at (5.05,-1.05)  {$c_2$};
  \node[orange!80!black] at (4.95,4.93)   {$c_3$};
  \node[orange!80!black] at (-0.95,4.93)  {$c_4$};
  \foreach \p in {(0,0),(4,0),(4,4),(0,4)} \fill \p circle (2.4pt);
  \node[violet!80!black] at (0.30,0.30) {\small $1$};
  \node[violet!80!black] at (3.70,0.30) {\small $2$};
  \node[violet!80!black] at (3.70,3.70) {\small $3$};
  \node[violet!80!black] at (0.30,3.70) {\small $4$};
  \node[blue!60!black] at (0.92,0.92) {\small $g_1$};
  \node[blue!60!black] at (3.08,0.92) {\small $g_2$};
  \node[blue!60!black] at (3.08,3.08) {\small $g_3$};
  \node[blue!60!black] at (0.92,3.08) {\small $g_4$};
  \node[blue!60!black] at (2,2.30) {core};
  \node[blue!60!black] at (2,1.72) {\small $(l_1,\dots,l_4,\ g_1,\dots,g_4)$};
\end{tikzpicture}
\caption{Components of a plaquette and its state.  The core (blue) is the data the plaquette operator changes, the
four active link states $l_1,\dots,l_4$ and the four site multiplicity indices
$g_1,\dots,g_4$.  The sixteen control links (orange, four per corner,
collectively $c_v$) do not change under the plaquette action but modulate
the matrix elements.  Each corner $v=1,\dots,4$ (violet) contains two active links, one multiplicity index, and four control links. The tensor product of the irreps on the six links (with conjugation for incoming links) must contain a singlet (the Gauss law) and the multiplicity index picks out which one. The link irreps
and the index together form the core data $s_v$ for the site. 
}
\label{fig:anatomy}
\end{figure}

Now let us discuss the plaquette operator and introduce some terminology (cf. Fig.~\ref{fig:anatomy}). The links of a plaquette are referred to as {\emph{active links}}, because their state can change under application of the plaquette operator. We use {\emph{corner}}, {\emph{vertex}}, and {\emph{site}} interchangeably to describe the lattice points where the plaquette's links meet. Links that are external to a plaquette but connect to its vertices are called {\emph{control links}}~\cite{ciavarella2021trailhead} for reasons that will become clear.  We define a plaquette (basis) state $s$ as four active link irreps $l_1,\dots,l_4$, four site singlet multiplicity indices $g_1,\dots,g_4$, and sixteen control link irreps. 

It will be   convenient to chop $s$ into different subsets of data. The active-link irreps $l_1,\dots,l_4$ we refer to as the {\emph{link core state}}. The link core state together with the multiplicity indices, $(\vec l, \vec g)$ we refer to simply as the {\emph{core state}}. At each site, the six link irreps and multiplicity index define the {\emph{site state}}. A site state at vertex $v$ separates into core data $s_v=(l_v,g_v)$ and control sector data $c_v$; the subscript will be used to denote data specific to a site of the plaquette. 

 An incoming plaquette state $s$ is referred to as a {\emph{source}}, a state $t$ connected to it by $\Box+\Box^\dagger$ is a {\emph{neighbor}}, and the transition $s\to t$ is called {\emph{forward}} when generated by $\Box$ and {\emph{reverse}} when generated by $\Box^\dagger$.  If $s\to t$ is generated by $\Box$, then it is never generated by $\Box^\dagger$.
  
The plaquette operator changes only core state.  The  matrix elements are a function of the state of the control links, but that data is unchanged by the operator.
 The local magnetic energy $\Box+\Box^\dagger$ connects a source state $s$ to its neighbors $t$ with matrix elements  $h_{st}=\langle t|H_p|s\rangle$.  $\Box$ has $105$ million nonzero matrix elements at $B=6$, billions at $B=9$, and they are not numbers that we can readily tabulate at higher truncations, so the primary task is to avoid implementing a single lookup table with one row per matrix element. That approach works well at lower dimensions and truncations~\cite{balaji2025circuits,balaji2025trunc,hidalgo2026lqcd} but it will not scale to full QCD near the continuum limit in $d=3$.  Instead we replace that table by a handful of much smaller ones. %

The matrix elements admit a per-vertex factorized form.  For physical plaquette states $s$ (source) and $t$ (target), we have
\begin{equation}
\langle t | \Box_p | s \rangle
=
\left(\prod_{k=1}^{4}\sqrt{\frac{\dim R^{s}_{\ell_k}}{\dim R^{t}_{\ell_k}}}\right)
\prod_{k=1}^{4} \mathcal{S}_{v_k}.
\label{eq:ME-master}
\end{equation}
Equation~\eqref{eq:ME-master} is built from an irrep-dimension prefactor over the four
plaquette links $\ell_k$ and four {\emph{site factors}}
$\mathcal{S}_{v_k}$, one per plaquette vertex.  Each site factor is a sum of products of Clebsch-Gordan coefficients,
\begin{equation}
\begin{aligned}
\mathcal{S}_{v} = \sum_{\sigma,r,r',u,u',\vec c} \phi(\sigma)\,
&\langle R'r' | (R,r)\otimes(f,\sigma)\rangle
\langle S'u' | (S,u)\otimes(\bar f,\tilde\sigma)\rangle \\
&\times
\langle \mathbf 1,g | (R,r)\otimes(S,u)\otimes(\vec C,\vec c)\rangle_F
\langle \mathbf 1,g' | (R',r')\otimes(S',u')\otimes(\vec C,\vec c)\rangle_F .
\end{aligned}
\label{eq:site-factor}
\end{equation}
Here the site factor is a function of $(R,S)$ and $(R',S')$, which are the  active link irreps at the vertex before/after the plaquette action; $\vec C$, the 
control link irreps which do not change; and $(g,g')$, the singlet multiplicity indices. $\phi$ is a conventional sign that generalizes Condon-Shortley phases. The subscript $F$ denotes the fixed (``F-order") convention for the spatial order in which the link irreps are slotted into the tensor product of the singlet Clebsch-Gordan coefficients. (In other words, we must fix a basis for our site singlets.)

The formula for the site factors above may appear somewhat hieroglyphic, but the content is physically simple. The site factors aggregate two wavefunctions (the initial and final states in the full electric basis, the second line of (\ref{eq:site-factor})), and two amplitudes (for the fundamental link operators to excite the two active links of the vertex into the final state, the first line of (\ref{eq:site-factor})). The summed indices in the four Clebsch-Gordan coefficients in (\ref{eq:site-factor}) are the internal color indices which are entirely integrated out in the gauge-invariant irrep basis. For a derivation and further details, see~\cite{balaji2025circuits,balaji2025trunc}.

The site factors can be classically precomputed and stored efficiently. We use the public code \texttt{pyclebsch} for their enumeration~\cite{balaji2025trunc}. This code returns a basis of real CGCs which are also organized into irreps of the permutation group acting on repeated irreps. The CGCs are  classically processed into site factors, and the site factors into sequences of rotation angles~\cite{Mottonen:2004dis} which can be loaded into the lookup tables of the block encoding below.

It will be convenient to treat the magnetic matrix element magnitudes and signs separately. We write the site factorization of the magnitudes more simply as
\begin{align}
|\Box_{st}|=w_1w_2w_3w_4,
\end{align}
 where $w_v\ge0$ is the magnitude of
the site factor of Eq.~\eqref{eq:site-factor} at vertex $v$, with a share of the irrep-dimension prefactor of Eq.~\eqref{eq:ME-master} folded into it.  (How the positive prefactor is shared among the vertices is arbitrary as long as it is fixed throughout the tables, and because it is positive it does not affect the sign.) We store the magnitudes $w_v$ and the sign $\sigma_v=\pm1$ of each site factor separately, and the full matrix element sign is $\sigma_{st}=\prod_v\sigma_v$.

Restricting attention to a single corner, the source data is
$s_v\equiv(\text{two link irreps},\,g)$ and a move sends it to
$t_v\equiv(\text{two new link irreps},\,g')$.  
A single link's irrep moves on a small part of the SU(3) fusion graph. 
As mentioned above, if there is a forward move  $s_v \underset{\Box}{\rightarrow} t_v$, then there is no forward move $t_v\underset{\Box}{\rightarrow} s_v$, but there is a reverse move $t_v\underset{\Box^\dagger}{\rightarrow} s_v$, and its matrix elements factorize into the same site factors,
 \begin{align}
 w^{\mathrm{fwd}}_v(s_v,c_v,t_v)&=w^{\mathrm{rev}}_v(t_v,c_v,s_v),\nonumber\\ 
 \sigma^{\mathrm{fwd}}_v(s_v,c_v,t_v)&=\sigma^{\mathrm{rev}}_v(t_v,c_v,s_v),
 \end{align}
 Therefore, only the forward site factors need be stored and we do not put a direction label on them, with the understanding that they are all  forward.  (Reverse moves will be handled by a bit which determines which slot of $w_v$ is read as the source and which as the target.)  Although $w_v$ itself carries no direction label, a source's forward neighbors (the $t_v$ reached by $s_v\underset{\Box}{\rightarrow} t_v$) form a different set from its reverse neighbors (the $t_v$ with $s_v\underset{\Box^\dagger}{\rightarrow} t_v$, equivalently $t_v\underset{\Box}{\rightarrow} s_v$), so the angle tables built from the $w_v$ in \S\ref{sec:oracle} are per-direction.

Let us conclude this section with some further comments on the basis choice and on the scope. 

\begin{itemize}
\item The original electric basis of Byrnes and Yamamoto~\cite{byrnes2006simulating} retains the full internal color content of each link, and therefore needs more qubits per link. The reduced electric basis used here keeps only an irrep label on each link and a singlet multiplicity index at each site.  Eliminating the internal color indices lowers the qubit count per link by a factor of a few, depending on how the original electric basis is encoded. A full accounting of the tradeoff must also weigh the cost of implementing the magnetic term, and explicit oracle constructions for the magnetic Hamiltonian within a block-encoding framework have not been carried out in the original electric basis.  Parts of the construction were described and asymptotic scaling formulas presented in~\cite{rhodes2024exponential}, but absent a complete construction the relative end-to-end complexity of the two approaches  has not been established.
\item The loop-string-hadron (LSH) basis~\cite{raychowdhury2020lsh,raychowdhury2020gausslaw} is another gauge-invariant reduction, reparameterizing the link and vertex data into local occupation numbers tied together by Abelian constraints.  In this approach vertex matrix elements become closed-form algebraic factors rather than tabulated recoupling data, and Gauss's law reduces to local number constraints.  It is developed furthest for SU(2); for SU(3) the formulation with dynamical quarks exists in one spatial dimension~\cite{kadam2023lsh_su3}, and the Hilbert space and operator representation of a trivalent SU(3) vertex have been constructed~\cite{kadam2025lsh_trivalent,kadam2025lsh_operator}. It would be of considerable interest to build a block encoding of the Hamiltonian in the LSH basis and compare query costs. 
\item The orbifold program~\cite{hanada2021orbifold,bergner2024orbifoldqcd,bergner2025orbifoldks,hanada2025gaugesym} replaces the compact group-valued links by non-compact bosonic variables whose truncation is a harmonic-oscillator level cutoff.  The exponential speedups reported there~\cite{bergner2025orbifoldks,halimeh2026universal,halimeh2025universalqcd,hanada2025bosons} refer to the dependence on that bosonic truncation level, which is different from the electric basis truncations we consider, and explicit $T$-counts for SU(3) simulations at continuum-level couplings have not yet been reported in the orbifold framework.  The two digitizations are therefore not yet comparable end to end, but again would be of considerable interest. 
\item Coupling the theory to dynamical quarks is beyond our scope, but the basic ingredients (site factors for Wilson and staggered fermions computed in the same irrep basis) were computed in~\cite{hidalgo2026lqcd}, and the oracles described here could be straightforwardly generalized to the link-matter operator. Also, in a similar spirit to the present work, Ref.~\cite{lamm2026chiral} developed a quantum signal processing implementation of overlap fermions. Since~\cite{lamm2026chiral} treats gauge dynamics as a black box, while the present work neglects matter, the two works are complementary.  An end-to-end simulation of QCD could combine the two. 
\end{itemize}

\section{Block Encoding Preliminaries}
\label{sec:blockencoding}

\subsection{Electric-magnetic decomposition}

The Kogut-Susskind Hamiltonian~(\ref{eq:KS-Hamiltonian}) splits into electric and magnetic terms. In the basis we consider the electric term is  relatively  straightforward to block encode and we relegate a brief discussion of a candidate encoding to Appendix~\ref{app:electric-be}. (The electric term might also be treated quite differently if one uses the interaction picture, or the hybrid Trotter-QSVT algorithm discussed below, which is why we do not commit to it.)  The magnetic term is more complicated and will be our focus.\footnote{The magnetic term of
Eq.~\eqref{eq:KS-Hamiltonian} is $\tfrac{1}{g^2}\sum_p(2N-\Box_p-\Box_p^\dagger)$.
The constant $2N$ per plaquette shifts the energy reference and is dropped, and
we take $H_B=\tfrac{1}{g^2}\sum_p(\Box_p+\Box_p^\dagger)$. The overall minus
sign could be applied by a $Z$ on the selector qubit inside $\mathrm{SELECT}$.}  Block encodings
of other Hamiltonians arising in high energy physics include~\cite{kane2025bosons,anderson2025krylov}.

\begin{figure}[htbp]
\centering
\begin{quantikz}[column sep=0.5cm, row sep={0.9cm,between origins}]
\lstick{$|0\rangle_{\rm sel}$}
  & \gate[2]{\mathrm{PREP}_{\rm LCU}} & \octrl{2} & \ctrl{1} & \gate[2]{\mathrm{PREP}_{\rm LCU}^\dagger} & \qw \\
\lstick{$|0\rangle_{p}$}
  & & \qw & \gate[2]{U_B} & & \qw \\
\lstick{$|\psi\rangle$}
  & \qw & \gate{U_E} & & \qw & \qw
\end{quantikz}
\caption{Block encoding of $H = H_E + H_B$ as an LCU. $p$ is a register whose basis states index each plaquette. The magnetic SELECT routes the single-plaquette oracle $U_p$ to the plaquette indexed by each $|p\rangle$ (Fig.~\ref{fig:ub-from-up}). }
\label{fig:lcu}
\end{figure}
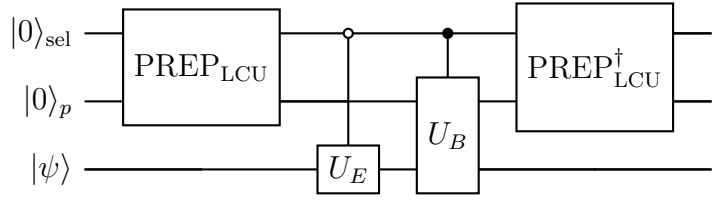

A block encoding is a unitary $U_H$
acting on the system plus a register $a$ of ancillas such that
\begin{equation}
\big(\langle 0|_a \otimes I\big)\, U_H\, \big(|0\rangle_a \otimes I\big)
= H/\beta ,
\end{equation}
where $\beta$ is the subnormalization. 
$H_E$ and $H_B$ can be block encoded separately, with subnormalizations $\alpha_E$ and $\alpha_B$, and 
combined in a linear combination of unitaries (LCU)~\cite{childs2012lcu},
\begin{align}
U_H=\mathrm{PREP}_{\rm LCU}^\dagger\,\mathrm{SELECT}\,\mathrm{PREP}_{\rm LCU},
\end{align}
via the circuit in Fig.~\ref{fig:lcu}. 
Here $\mathrm{PREP}_{\rm LCU}$ creates the state
\begin{equation}
\sqrt{\frac{\alpha_E}{\beta}}\,|0\rangle_{\rm sel}|0\rangle_p
+\sqrt{\frac{\alpha_B}{\beta}}\,|1\rangle_{\rm sel}
\otimes\frac{1}{\sqrt{N_{\rm plaq}}}\sum_p|p\rangle,
\label{eq:prep-lcu}
\end{equation}
using one $R_y$ %
on the
selector qubit and a uniform preparation of the plaquette index register $p$. Both $sel$ and $p$ are a part of the block encoding ancilla register $a$.
In the SELECT part, $U_B$ reads the
plaquette register $p$ and applies the single-plaquette oracle $U_p$ to each plaquette in parallel. This relies on a routing circuit to swap
each plaquette's subregister to the oracle's  position,
\begin{equation}
U_B=\sum_p |p\rangle\langle p|_p\otimes \mathrm{Route}(p)^\dagger\,U_p\,\mathrm{Route}(p)
\label{eq:ub-from-up}
\end{equation}
or as a circuit,
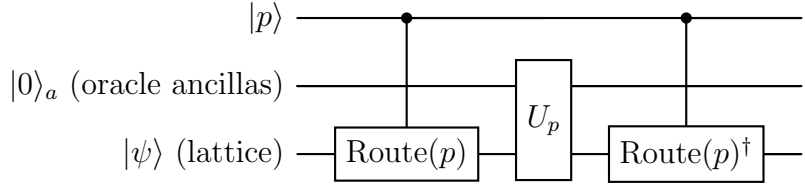
\begin{figure}[htbp]
\centering
\begin{quantikz}[column sep=0.5cm, row sep={0.9cm,between origins}]
\lstick{$|p\rangle$}
  & \ctrl{2} & \qw & \ctrl{2} & \qw \\
\lstick{$|0\rangle_{a}$ (oracle ancillas)}
  & \qw & \gate[2]{U_p} & \qw & \qw \\
\lstick{$|\psi\rangle$ (lattice)}
  & \gate{\mathrm{Route}(p)} & & \gate{\mathrm{Route}(p)^{\dagger}} & \qw
\end{quantikz}
\caption{The whole-lattice magnetic block encoding $U_B$ assembled from the
single-plaquette oracle $U_p$,
embedded as 
the magnetic SELECT branch in Fig.~\ref{fig:lcu}. }
\label{fig:ub-from-up}
\end{figure}

The circuit in Fig~(\ref{fig:lcu}) block encodes $H$ with lattice coupling $g=1$ and $\beta = \alpha_E + \alpha_B$.\footnote{$g=1$ is a reasonable first ``near continuum limit" target.} 
The magnetic subnormalization is $\alpha_B = N_{\rm plaq}\,\alpha_p$, where $\alpha_p$ is the single-plaquette subnormalization that depends on the truncation. At $B=9$ we find $\alpha_p\approx 56$ and  $\alpha_B$ is greater than the electric subnormalization $\alpha_E = \tfrac{1}{2}N_{\rm links}\,C_2^{\max}$. $C_2^{\max}=10/3$ for the active irrep set that includes up to the two-index tensors. Both $\alpha_E$ and $\alpha_B$ are extensive, and the size-independent comparison is the per-link electric weight $\tfrac{1}{2}C_2^{\max}=\tfrac53\ll 56$.

For other values of $g$, we have the overall subnormalization
\begin{equation}
\beta(g)=g^2\,\alpha_E+\frac{\alpha_B}{g^2},
\label{eq:beta-of-g}
\end{equation}
with $\alpha_E$ and $\alpha_B$ given by the $g=1$ values above.  In the circuit,  $g$ can be absorbed into the $R_y$ that $\mathrm{PREP}_{\rm LCU}$ applies to the selector qubit in Eq.~\eqref{eq:prep-lcu}.

Let us trace the  circuit at the current level of granularity. Starting in $|0\rangle_{\rm sel}|0\rangle_p|\psi\rangle$, $\mathrm{PREP}_{\rm LCU}$ creates the state of Eq.~\eqref{eq:prep-lcu}, tensored with $|\psi\rangle$. $\mathrm{SELECT}$ then applies $U_E$ to $|\psi\rangle$ on the $|0\rangle_{\rm sel}$. On the $|1\rangle_{\rm sel}$ branch, it creates
 $\sqrt{\tfrac{\alpha_B}{\beta}}\,\tfrac{1}{\sqrt{N_{\rm plaq}}}\sum_p|1\rangle_{\rm sel}|p\rangle\,U_B^{(p)}|\psi\rangle$. Finally $\mathrm{PREP}_{\rm LCU}^\dagger$ inverts the uniform preparation of $p$ and the selector rotation. On the $p$ register $\mathrm{PREP}_{\rm LCU}$ acts as $\tfrac{1}{\sqrt{N_{\rm plaq}}}\sum_p\langle p|$ (and the $\langle0|_{\rm sel}$ projection supplies a second $\sqrt{\alpha_B/\beta}$), so keeping only the $|0\rangle_{\rm sel}|0\rangle_p$ component we obtain $\tfrac{\alpha_B}{\beta}\tfrac{1}{N_{\rm plaq}}\sum_p U_B^{(p)}|\psi\rangle$. 
In total, projecting onto the block-encoding corner leaves $\tfrac{\alpha_E}{\beta}\tfrac{H_E}{\alpha_E}+\tfrac{\alpha_B}{\beta}\tfrac{1}{N_{\rm plaq}}\sum_p\tfrac{H_p}{\alpha_p}=\beta^{-1}(H_E+H_B)=H/\beta$ on $|\psi\rangle$, using $\alpha_B=N_{\rm plaq}\alpha_p$.

Our main task is to block encode $\Box_p+\Box_p^\dagger$ which forms the building block of the magnetic Hamiltonian. First, however, we review the lookup table and programmed rotation primitives that appear throughout the algorithm.

\subsection{Quantum lookups and programmed rotations}
\label{sec:data-quantum}

\subsubsection{Lookups}
A quantum lookup is the analog of a classical read-only table~\cite{babbush2018qrom,low2019dirty,berry2019qubitization}.
Given a classical table $D$ with $N$ $m$-bit entries, a lookup is a unitary operator acting on a key register $|k\rangle$ and a
clean  output register $|0\rangle_m$ that acts as: 
\begin{equation}
D\,|k\rangle\,|0\rangle_m=|k\rangle\,|D(k)\rangle_m,
\qquad k\in\{0,\dots,N-1\}.
\label{eq:qroam-def}
\end{equation}
Passing a superposition of keys to the lookup loads all the  data in parallel.

Various types of quantum lookups differ in how the incoming data is read. For tables whose listed keys occupy only a sparse subset of the key register, it is convenient and simple to use sparse unary
iteration~\cite{babbush2018qrom,gidney2018halving}, explained in detail in App.~\ref{app:sparse-qrom}. This may not be the most optimal approach.  Hashing the sparse keys followed by a dense  lookup may be better. We leave such lookup optimizations as targets for future work, tabulated with other optimization possibilities in~\S\ref{sec:conclusions}. The $T$-gate cost of calling QROM with the sparse unary iteration lookup is $4(N-1)\,T$ to load and the same to erase. 

\subsubsection{Programmed Rotations}
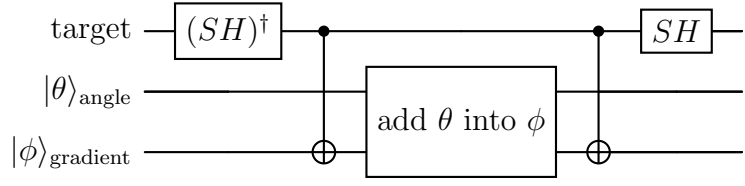
\begin{figure}[htbp]
\centering
\begin{quantikz}[column sep=0.4cm, row sep={0.8cm,between origins}]
\lstick{target}
  & \gate{(SH)^{\dagger}} & \ctrl{2} & & \ctrl{2}  & \gate{SH} & \qw \\
\lstick{$|\theta\rangle_{\rm angle}$}
  & \qw && \gate[2]{\text{add}\ \theta\ \text{into}\ \phi} & & \qw & \qw \\
\lstick{$|\phi\rangle_{\rm gradient}$}
  && \targ{} &&\targ{}& \qw & \qw
\end{quantikz}
\caption{A programmed rotation $R_y(2\theta)$ on the target wire.}
\label{fig:prog-rot}
\end{figure}

We use lookups to store and recall the angles of rotation trees. Since the angles $\theta$ are encoded in $b_{\rm rot}$-qubit  register amplitudes rather than hard-coded, the rotations are ``programmed," and a single rotation is implemented as follows (cf. Fig.~\ref{fig:prog-rot}). 

Let the angle register hold the integer $a$, encoding $\theta=2\pi a/2^{b_{\rm rot}}$, and define $\omega=e^{2\pi i/2^{b_{\rm rot}}}$. $a\rightarrow  2^{b_{\rm rot}}-a$ is a reflection $\theta\rightarrow 2\pi-\theta$, which is a convenient property. An ancilla holds a phase-gradient resource state,  $|\phi\rangle_{\rm gradient}=2^{-b_{\rm rot}/2}\sum_k\omega^{k}|k\rangle$~\cite{gidney2018halving}. (The resource state is prepared once and reused.)  Conditioned on the target qubit, the angle register is added into the gradient register. The conditioning is such that on the target's $|1\rangle$ branch the modular addition shifts $k$ by $- a$, and on the target's $|0\rangle$ branch it shifts $k$ by $a$.  Upon relabeling the summation index we return to the same state times a phase,
\begin{align}
|0\rangle|a\rangle|\phi\rangle 
&\rightarrow
|0\rangle |a\rangle2^{-b_{\rm rot}/2}\sum_k\omega^{k}|k+a\rangle
=\omega^{-a} |0\rangle |a\rangle |\phi\rangle\nonumber\\
|1\rangle |a\rangle |\phi\rangle
&\xrightarrow{ \rm CX }
|1\rangle |a\rangle 2^{-b_{\rm rot}/2}\sum_k\omega^{k}|2^{b_{\rm rot}}-1-k\rangle
\nonumber\\
&~=~|1\rangle |a\rangle 2^{-b_{\rm rot}/2}\sum_k\omega^{-k-1}|k\rangle\nonumber\\
&\xrightarrow{+a}
|1\rangle|a\rangle2^{-b_{\rm rot}/2}\sum_k\omega^{-k-1}|k+a\rangle\nonumber\\
&\xrightarrow{\rm CX}|1\rangle|a\rangle2^{-b_{\rm rot}/2}\sum_k\omega^{-k-1}|2^{b_{\rm rot}}-1-k-a\rangle\nonumber\\
&~=~|1\rangle |a\rangle 2^{-b_{\rm rot}/2}\sum_k\omega^{a}|k\rangle\nonumber\\
&~=~\omega^{a}|1\rangle\,|a\rangle\,|\phi\rangle ,
\label{eq:kickback}
\end{align}
Thus the  effect on the target is  phase kickback by $R_z(2\theta)$.  Conjugating by $SH$ gives the desired rotation, 
\begin{equation}
(SH)\,R_z(2\theta)\,(SH)^{\dagger}\,|0\rangle
=R_y(2\theta)\,|0\rangle=\cos\theta\,|0\rangle+\sin\theta\,|1\rangle .
\label{eq:clifford-sandwich}
\end{equation}

The  non-Clifford cost of the programmed rotations is in the addition step, about one Toffoli per angle bit, so each  costs $4\,b_{\rm rot}\,T$.

\subsubsection{Programmed Rotation Trees}
Generally the state we want to prepare is a superposition over a multi-qubit register. This state can be prepared by a binary $R_y$ tree, which is parametrized by a collection of angles, one angle per node of the tree. In the  tables we store each key's full angle collection in one row, using $m=(\text{tree angles})\times b_{\rm rot}$ bits. The lookups  then return the key's tree angles in a fixed node order, and  the rotations are performed in that order.  The tree  acts on an
$n$-qubit target as
\begin{equation}
|0\cdots0\rangle\mapsto\sum_{x_1\cdots x_n}
\prod_{j=1}^{n}
\big(\cos\theta_{x_1\cdots x_{j-1}}\big)^{1-x_j}
\big(\sin\theta_{x_1\cdots x_{j-1}}\big)^{x_j}
|x_1\cdots x_n\rangle.
\label{eq:ry-tree}
\end{equation}
 The first rotation
$R_y(2\theta_{\emptyset})$ acts on the leading qubit of the target register, and each later rotation $R_y(2\theta_x)$ acts on the next qubit conditioned on the bitstring $x$  written into the earlier qubits~\cite{Mottonen:2004dis}. 

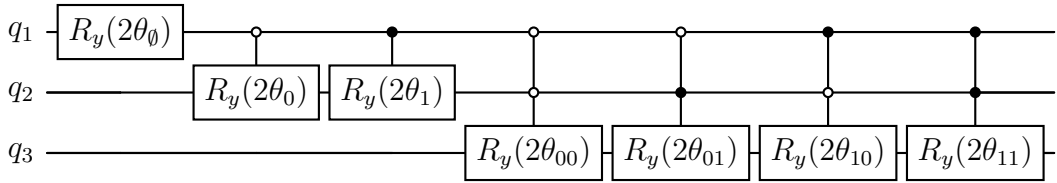
\begin{figure}[htbp]
\centering
\begin{quantikz}[column sep=0.14cm, row sep={0.8cm,between origins}]
\lstick{$q_1$} & \gate{R_y(2\theta_{\emptyset})} & \octrl{1} & \ctrl{1} & \octrl{1} & \octrl{1} & \ctrl{1} & \ctrl{1} & \qw \\
\lstick{$q_2$} & \qw & \gate{R_y(2\theta_{0})} & \gate{R_y(2\theta_{1})} & \octrl{1} & \ctrl{1} & \octrl{1} & \ctrl{1} & \qw \\
\lstick{$q_3$} & \qw & \qw & \qw & \gate{R_y(2\theta_{00})} & \gate{R_y(2\theta_{01})} & \gate{R_y(2\theta_{10})} & \gate{R_y(2\theta_{11})} & \qw
\end{quantikz}
\caption{A three-qubit rotation tree.  Each box is a programmed rotation of Fig.~\ref{fig:prog-rot}, and the wires from earlier qubits are controls on the phase-gradient addition.}
\label{fig:ry-tree-ex}
\end{figure}

In a tree of programmed rotations,  the node-$x$ rotation becomes the programmed rotation of Fig.~\ref{fig:prog-rot}, with the condition that the previous target qubits equal $x$. The CX layers of Fig.~\ref{fig:prog-rot} are also added in. Fig.~\ref{fig:ry-tree-ex} shows a three-qubit example. Thus on any branch of the wavefunction only one rotation per tree level fires.  A preparation over $k$ basis states is parametrized by at most $k-1$ tree angles. (A subtree carrying no weight receives angle zero.)  The tree controls add to the $T$ cost, one Toffoli for each control, plus one per angle bit to condition the addition.

\subsubsection{Generalized Programmed Swap Networks}
A rotation tree with $k$ branches corresponds to superposition state on $n=\lceil\log_2 k\rceil$ qubits.  When the superposition of physical states that we want to prepare is fairly dense, the rotation tree can act on the target register directly. When instead we want to populate a subspace with a sparse superposition, we use a different implementation, and include additional data in the lookup table for the tree. 

First, the lookup table needs to return not only the tree angles for each key, but also the reachable target states. So the tables store angle data in a fixed order, and also target data, ordered in the same way as the branches of the rotation tree. Slot $j$ of the target data in the row is the bitstring corresponding to  branch $j$'s target state, $v_j$,  in the same encoding as used by the target register. When a key is passed to the lookup, its angle list is printed into an angle register, and its target list is printed into a value register. The value register holds one slot per branch, each as wide as the target register, so for $k$ branches it is $k$ times the target register's size (for the proposal stage of \S\ref{sec:stagea} at $B=9$, this is 32 slots $\times$ 13 qubits $=$ 416 qubits).

Next, we use the programmed rotation tree to prepare the superposition over the states of a small work register, numbered  $u=0\dots k-1$. To move this superposition into the actual target register requires a separate  circuit $W$, implementing an isometry from the $k$-dimensional computational subspace of $u$ onto the $k$-dimensional subspace of the target register spanned by $|v_0\rangle,\dots,|v_{k-1}\rangle$ and returning $u$ to
$|0\rangle$:
\[
W\,\sum_{j=0}^{k-1}\alpha_j\,|j\rangle_u\,|0\rangle_{\rm tgt}
=\sum_{j=0}^{k-1}\alpha_j\,|0\rangle_u\,|v_j\rangle_{\rm tgt}.
\]
Thus $W$  swaps the logical content of a small register into a subspace of a larger one. Furthermore  the swap map inside $W$ is specified by data in the value register at run time, not by a fixed map compiled into gates. So $W$ is to a swap network what a programmed rotation is to a compiled rotation, and in this sense we may think of $W$ as a ``programmed swap network."\footnote{More precisely, $W$ as written here is the one-directional half of a swap, defined on a  subspace in which $u$ is guaranteed to hold $0,\dots,k-1$ and the target register is guaranteed to be in $|0\rangle$.  On that subspace it transfers $|j\rangle|0\rangle\to|0\rangle|v_j\rangle$.  Off the  subspace the circuit acts as garbage.  A true two-sided SWAP would also need $|0\rangle|v_j\rangle\to|j\rangle|0\rangle$ in the same circuit. For this reason we call it a ``generalized" programmed swap network.}   The same conversion from a dense counter to a sparse pattern appears inside the sparse state preparation of~\cite{berry2019qubitization}.

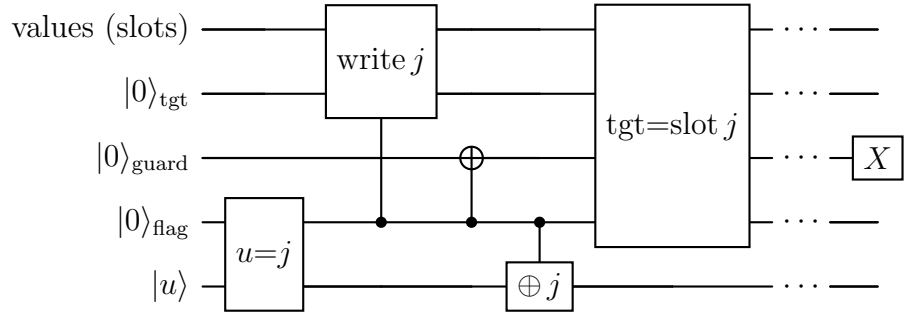
\begin{figure}[htbp]
\centering
\begin{quantikz}[column sep=0.3cm, row sep={0.85cm,between origins}]
\lstick{values (slots)}
  & \qw & \gate[2]{\mathrm{write}\,j} & \qw & \qw & \gate[4]{\mathrm{tgt}{=}\mathrm{slot}\,j} & \ \cdots\  & \qw \\
\lstick{$|0\rangle_{\rm tgt}$}
  & \qw &  & \qw & \qw &  & \ \cdots\  & \qw \\
\lstick{$|0\rangle_{\rm guard}$}
  & \qw & \qw & \targ{} & \qw &  & \ \cdots\  & \gate{X} \\
\lstick{$|0\rangle_{\rm flag}$}
  & \gate[2]{u{=}j} & \ctrl{-2} & \ctrl{-1} & \ctrl{1} &  & \ \cdots\  & \qw \\
\lstick{$|u\rangle$}
  &  & \qw & \qw & \gate{\oplus\,j} & \qw & \ \cdots\  & \qw
\end{quantikz}
\caption{The write circuit $W$, drawn with one representative block.  See
 text for the definition of each gate.}
\label{fig:write-w}
\end{figure}

The circuit that implements $W$ is shown in Fig.~\ref{fig:write-w}.  It contains one block for each branch number $j=0,\dots,k-1$, run in ascending order and laid out in the sequence denoted by ellipses.  First, a multi-controlled X gate  computes $u=j$ into a block flag.\footnote{In usage below, one can also add a control requiring that a reject flag $\mathrm{F}$ is unset. This prevents the reject branch, on which the $u$ register is empty, from firing on block $0$. Strictly this is not needed for correctness, since the final projection of the block encoding discards that branch either way.} 
Next, the value stored in slot $j$ of the value register is written into the target register, one bit at a time, by Toffolis controlled on the  flag and each bit of the value register. The  flag is also used to set one guard qubit appended to the target register. Afterward we uncompute the branch register by CX gates.  The  flag is uncomputed by testing ``target register $=v_j$ and guard bit $=1$."\footnote{Without the guard, the test would misfire on incorrect branches of the wavefunction whenever $v_j$ is itself the all-zero state, which is allowed by the encoding. The test would see $v_j=$ target register for branches that have not yet been reached, where the target register is also the all-zero state, and incorrectly flip the flag bit from zero to one. The ascending order is important for a similar reason: a branch that has finished its turn in $W$ has $u=0$ afterward, which would trigger the $u=j$ gate if $j=0$ was not the first block. With ascending order each branch is matched in exactly one block.}

After the last block, the guard qubit is in the $|1\rangle$ state on every branch, and a single X returns it to zero.

\subsubsection{Lookup + Programmed rotation blocks}
\begin{figure}[htbp]
\centering
\begin{quantikz}[column sep=0.3cm, row sep={0.8cm,between origins}]
\lstick{key}
  & \gate[2]{\mathrm{lookup}(N,\,m)} & \qw & \gate[2]{\mathrm{lookup}^{\dagger}} & \qw \\
\lstick{$|0\rangle_{\rm angle}$}
  & & \gate[2]{R_y\ \text{tree}} & & \rstick{$|0\rangle$}\qw \\
\lstick{target}
  & \qw & & \qw & \qw
\end{quantikz}
\caption{The simpler version of the lookup-rotation circuit block.}
\label{fig:stage-shape}
\end{figure}
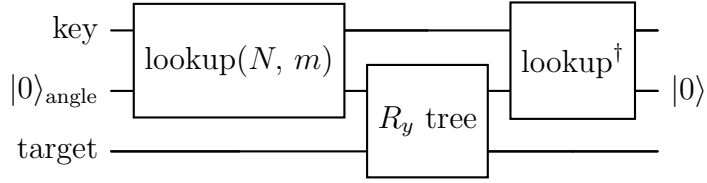

\begin{figure}[htbp]
\centering
\begin{quantikz}[column sep=0.3cm, row sep={0.8cm,between origins}]
\lstick{key}
  & \gate[2]{\mathrm{lookup}(N,\,m)} & \qw & \qw & \gate[2]{\mathrm{lookup}^{\dagger}} & \qw \\
\lstick{$|0\rangle$ (angles, values)}
  & & \gate[2]{R_y\ \text{tree}} & \gate[3]{W} & & \rstick{$|0\rangle$}\qw \\
\lstick{$|0\rangle_{u}$}
  & \qw & & & \qw & \rstick{$|0\rangle$}\qw \\
\lstick{$|0\rangle_{\rm tgt}$ (guard)}
  & \qw & \qw & & \qw & \qw
\end{quantikz}
\caption{The sparse-target version of the lookup-rotation block.  The tree prepares the branches on $u$, and the
write step $W$ moves them onto the target register.}
\label{fig:stage-shape-sparse}
\end{figure}
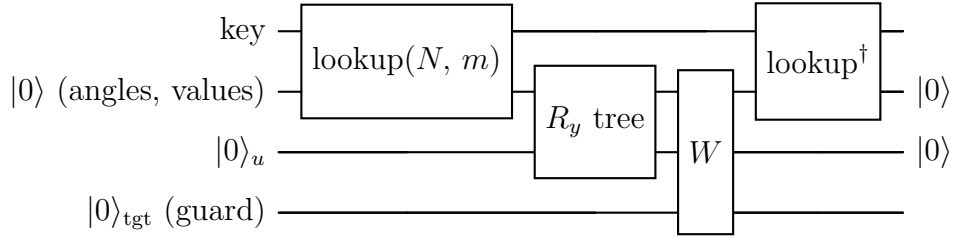

Putting the pieces of the previous sections together gives two versions of the basic circuit block that appears throughout the oracles.  Both are of lookup + programmed rotation form. In the first version, the superposition to be prepared is dense enough that the rotation tree can act on the target register directly. This version is shown in
Fig.~\ref{fig:stage-shape}.
In the second version, where the target superposition is sparse, we must use the intermediate stage with the programmed swap network.  
This version is shown in
Fig.~\ref{fig:stage-shape-sparse}. 

\section{Overview of the single-plaquette oracle}
\label{sec:spo}
\begin{figure}[t!]
\centering
\begin{quantikz}[column sep=0.3cm, row sep={0.85cm,between origins}]
\lstick{$|0\rangle_{\rm DIR}$}
  & \gate[4]{\mathrm{PREP}}
  &  \gate{X}
  & \gate[4]{\mathrm{PREP}'^{\dagger}}
  & \rstick{$\langle0|$}\qw \\
\lstick{$|0\rangle_{\rm F}$}
  &  & \qw
  & & \rstick{$\langle 0|$}\qw \\
\lstick{$|0\rangle_{\rm TGT}$}
  & & \swap{1} 
  & & \rstick{$\langle 0|$}\qw \\
\lstick{$|s\rangle_{\rm SRC}$}
  & & \targX{} 
  & & \rstick{$\langle t|$}\qw
\end{quantikz}
\caption{The single-plaquette magnetic oracle $U_p$ which block-encodes $H_p/\alpha_p$.  PREP and $\mathrm{PREP}'^{\dagger}$  are the  expensive gates. The swap, which exchanges
TGT with the core fields of SRC, and the $X$ on DIR, are Clifford gates. The matrix element sign is carried by the signed corner angles of PREP.  SRC is a subregister of the lattice data register $|\psi\rangle$ and holds the plaquette state.  DIR, F, TGT (and the tag and key registers, ${\rm TAG}$ and KEY, introduced below) are part of the block-encoding ancillas in the register $a$. The bra notation on the right-hand side denotes the projection that defines the block encoded matrix element $\langle t,0|\, U_p\,|s,0\rangle$. (In later figures ket notation on the right hand side denotes states that emerge unitarily.)}
\label{fig:oracle}
\end{figure}
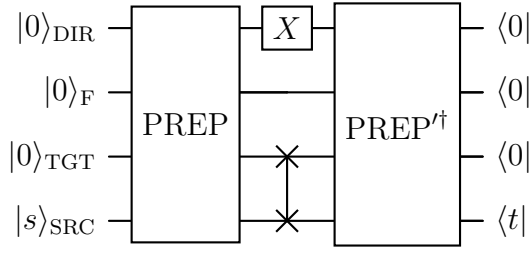

As discussed above, the magnetic block encoding is built from a single-plaquette magnetic oracle $U_p$ which block-encodes $H_p/\alpha_p$.  We now construct this oracle, which is shown at a high level in Fig.~\ref{fig:oracle}. SRC is the subregister of the lattice register $|\psi\rangle$ holding  the plaquette state. Other registers are part of the block encoding ancilla collection. 

We will explain the circuit elements and the roles of the ancilla registers in subsequent sections, but in brief:
\begin{itemize}
\item \textbf{PREP} is keyed on the plaquette source state in the lattice data subregister SRC. It rotates block encoding ancilla registers DIR, TGT, and F into a weighted neighbor superposition, with amplitude $\sigma_{st}\sqrt{|h_{st}|/\alpha_p}$ on the branch carrying neighbor $t$'s core and direction bit.  The matrix element sign $\sigma_{st}$ is folded into the corner angles of this unitary.
\item \textbf{SWAP} (\S\ref{sec:swap}) exchanges the  target core in TGT with the core fields of SRC, leaving the unchanged control links in place.  This is what will capture the off-diagonal nature of the block encoded $\Box+\Box^\dagger$ in the electric basis. SRC now holds the neighbor, and TGT holds the original source core.
\item \textbf{X(DIR)} flips the direction qubit: an edge that is forward for the source is a reverse edge for the neighbor, and after the swap the direction
must be flipped.
\item $\mathbf{PREP'^{\dagger}}$ is keyed on
the neighbor state held in SRC.  It is the circuit adjoint of  $\mathrm{PREP}'$, a unitary that stores the unsigned corner angles rather
than the signed ones of PREP. In this way the two prepares together apply the
matrix element sign  once.
\end{itemize}
Projecting the ancillas DIR, F, and TGT onto $|0\rangle$, $U_p$ block-encodes $\Box+\Box^\dagger$  with
subnormalization $\alpha_p$.  Routing then applies $U_p$ across all $N_{\rm plaq}$ plaquettes in superposition to form the whole-lattice $U_B$.

Now let us discuss the
single-plaquette subnormalization $\alpha_p$. 
Given an incoming plaquette source state $|s\rangle$, the unsigned oracle $\mathrm{PREP}'$ prepares the forward and reverse neighbors of $s$ with amplitudes
proportional to the square roots of the matrix element magnitudes,
$\sqrt{|h_{st}|}$. 
We define the  row-$1$-norm of row $s$ as
\begin{equation}
C_s=\sum_t|h_{st}|. 
\label{eq:row1norms}
\end{equation}
With the amplitude-weighted prepare, the single-plaquette subnormalization is bounded below by the maximum row-$1$-norm,
\begin{equation}
\alpha_p \ge C = \max_s C_s \equiv
\text{the maximum row-$1$-norm of } H_p .
\label{eq:row1norm}
\end{equation}
This is a unitarity bound: $\sum_t\sqrt{|h_{st}|/\alpha_p}\,|t\rangle$ has squared norm $\sum_t|h_{st}|/\alpha_p=C_s/\alpha_p$, and demanding $C_s/\alpha_p\le1$ for every source requires  $\alpha_p\ge\max_s C_s=C$.\footnote{The reason for a single subnormalization for all sources is that in the full circuit the prepare
and the unprepare are keyed on different rows ($s$ and $t$), and their two
amplitude factors multiply.} 

Since PREP prepares amplitudes $\sigma_{st}\sqrt{|h_{st}|/\alpha_p}$, each source must store its unused amplitude $\sqrt{1-C_s/\alpha_p}$ on a flagged ``reject'' branch. Because the block encoding post-selects every ancilla onto $|0\rangle$,  the reject branch is discarded.

%

 $\alpha_p$ sits well above the spectral norm $\|H_p\|$. $H_p$ is block-diagonal in the control links, so $\|H_p\|$ is the maximum eigenvalue over control sectors, attained in the vacuum sector at both truncations.  At $B=6$ $\|H_p\|\approx 2.6$,  while at $B=9$ $\|H_p\|\approx 3.2$.  The matrix elements in a heavy row carry mixed signs that cancel in the eigenvalue while adding in the row sum, and our PREP does not benefit from that cancellation.

Next we turn to the explicit construction. \S\ref{sec:oracle} constructs PREP, and \S\ref{sec:finishing} assembles  the full block encoding $U_p$.

\section{PREP oracle}
\label{sec:oracle}

The most complicated part of the circuit is the amplitude-weighted PREP oracle.  Given a source state, PREP rotates the ancilla registers TGT, DIR, and F
into a  superposition over all the neighbors $t$ of $s$:
\begin{equation}
\begin{aligned}
\mathrm{PREP}|s\rangle_{\rm SRC}\,&|\mathrm{0}\rangle_{\rm TGT}|0\rangle_{\rm DIR}|0\rangle_{a}\\
&=
|s\rangle_{\rm SRC}\Big[\frac{1}{\sqrt{\alpha_p}}\sum_{t}\sigma_{st}\sqrt{|h_{st}|}
|\mathrm{core}(t)\rangle_{\rm TGT}|d_t\rangle_{\rm DIR}|0\rangle_{a}
+\sqrt{1-C_s/\alpha_p}|\bot\rangle\Big].
\end{aligned}
\label{eq:weighted-prep}
\end{equation}

$|\mathrm{core}(t)\rangle$ is the core state (active link irreps + site factor multiplicities) of each neighbor, written into TGT. As above, $C_s=\sum_t|h_{st}|$ is the row-$1$-norm
of source $s$, $\alpha_p\ge \max_s C_s$ the single-plaquette subnormalization of Eq.~\eqref{eq:row1norm}, and $\sigma_{st}=\operatorname{sign}h_{st}$ is the matrix element sign. Forward neighbors, reached by the application of $\Box$, have direction bit $d_t=0$.  Reverse neighbors, reached by the application of $\Box^\dagger$, have $d_t=1$.   The ancilla register $a$ contains  rejection flags and their tags, as  described in \S\ref{sec:stages}.  $|\bot\rangle$ is a normalized state of TGT, DIR and $a$ with no support in the $|0\rangle_a$ sector: every rejected configuration carries at least one flag set, so it is orthogonal to the accepted branch on the ancilla state, regardless of what TGT and DIR hold.  Its precise form is unimportant.


Directly implementing Eq.~\eqref{eq:weighted-prep} with a single lookup is infeasible. The distinct rows  number in the tens of millions already in the example truncations we consider, and the matrix elements number in the billions.  The factorization $|\Box_{st}|=w_1w_2w_3w_4$ of Eq.~\eqref{eq:ME-master} is the obvious structure to exploit, but the four site factors are correlated by the active links of the plaquette.  Adjacent corners share an active link and must agree on its new value, and a combination of individually allowed corner moves generally does not correspond to a valid plaquette transition.

There are different approaches to implementing these correlations. One could walk the plaquette vertices one at a time, demanding consistency with the predecessors. We will take a different path that implements the correlation first.  Our PREP circuit  begins by proposing a target state for the active link core and a direction, $(\lambda',\mathrm{DIR})$, from a table keyed on the   source link core state. Then four corner stages rotate amplitude in and out of the reject branch and the multiplicity index registers for each corner, depending on the proposed $(\lambda',\mathrm{DIR})$ and the state of the control links.

\begin{figure}[htbp]
\centering
\begin{quantikz}[column sep=0.22cm, row sep={0.72cm,between origins}]
\lstick{$|s\rangle_{\rm SRC}$}
  & \gate[2]{\mathrm{ADDR}} & \gate[5]{\substack{\text{A}\\[1pt](\lambda',\,\mathrm{DIR},\,\mathrm{F})}} & \gate[7]{B_1} & \gate[7]{B_2} & \gate[7]{B_3} & \gate[7]{B_4} & \gate[2]{\mathrm{ADDR}^{\dagger}} & \rstick{$|s\rangle$}\qw \\
\lstick{$|0\rangle_{\rm KEY}$}
  & & \linethrough & & & & & & \rstick{$|0\rangle$}\qw \\
\lstick{$|0\rangle_{\rm TGT.LNK}$}
  & \qw & & & & & & \qw & \qw \\
\lstick{$|0\rangle_{\rm DIR}$}
  & \qw & & & & & & \qw & \qw \\
\lstick{$|0\rangle_{\rm F}$}
  & \qw & & & & & & \qw & \qw \\
\lstick{$|0\rangle_{\rm TGT.G}$}
  & \qw & \qw & & & & & \qw & \qw \\
\lstick{$|0\rangle_{{\rm TAG}}$}
  & \qw & \qw & & & & & \qw & \qw
\end{quantikz}
\caption{The PREP circuit realizing Eq.~\eqref{eq:weighted-prep}.  ADDR is the class lookup, which reads each corner's environment from SRC and writes the four
compact class labels into KEY.  The proposal stage (A) is keyed on the source link core state, and it writes the four target links and the direction into TGT.LNK and DIR and sets the proposal flag F.  The four corner stages ($B_1,\dots,B_4$) write site singlet multiplicity indices into TGT.G or set the corner reject flags, keyed on SRC, KEY, TGT.LNK, and DIR.
$\mathrm{ADDR}^{\dagger}$ returns KEY to $|0\rangle$.  The tag register ${\rm TAG}$ records the direction under which each rejected branch was vetoed.}
\label{fig:prep-circuit}
\end{figure}
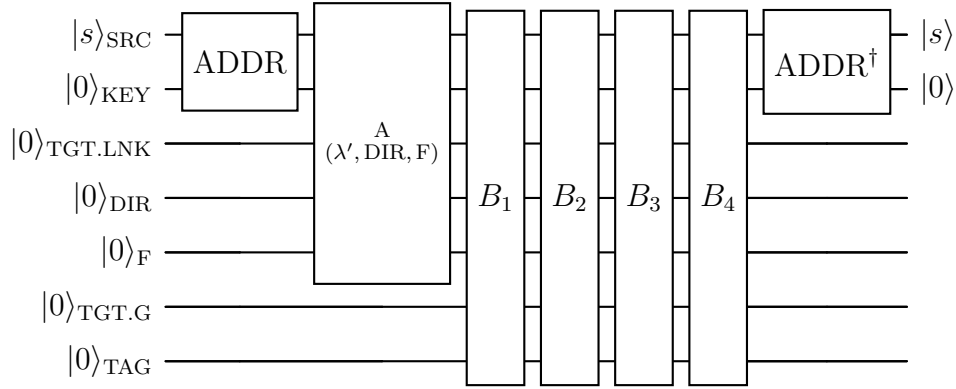

The circuit is shown at a high level in Fig.~\ref{fig:prep-circuit}. We discuss each subcomponent in the following subsections.

In Fig.~\ref{fig:prep-circuit}  we introduce two additional registers beyond those of Fig.~\ref{fig:oracle} and divide the target core register TGT into two subregisters.  TGT.LNK will the four target link states $(l_1',\dots,l_4')$ populated in the proposal stage. TGT.G will hold the four site singlet multiplicity indices $(g_1',\dots,g_4')$, prepared  during the corner stages.  Each TGT.G slot also includes a one-qubit reject flag $\mathrm{F}_v$, the analog for corner $v$ of the proposal flag F. A set reject flag indicates a failed preparation at that stage.  KEY is a work register into which $\mathrm{ADDR}$ will copy the four class labels defined below.  ${\rm TAG}$ holds five one-bit direction tags, one for each of the five flags; its role in the  oracle is explained in \S\ref{sec:correctness}.

\subsection{Control classes and the class lookup}
\label{sec:address}
ADDR is an equivalence class lookup run at the beginning of PREP.  It is realized by a QROM implemented with sparse unary iteration (App.~\ref{app:sparse-qrom}). ADDR's job is to read each corner's raw environment and return a compact class label that keys the later corner stages, and its purpose is data compression to save on rows of the corner tables. Although for the most part we do not attempt to optimize the circuit construction in this paper, we do attempt to use the physics of the lattice gauge theory to optimize the algorithm (which is the whole point of the construction.) ADDR is an example. It is not essential to the block encoding, but it does illustrate how specific structures in the matrix elements can be used to implement the transitions more efficiently.

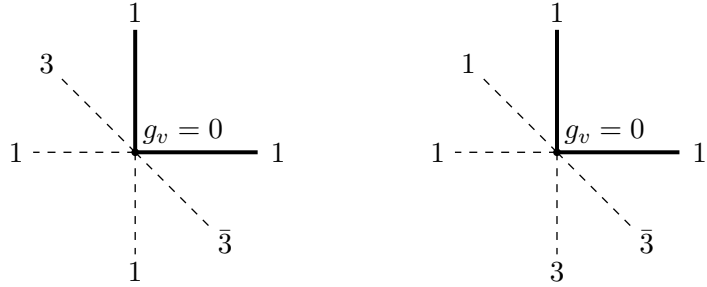
\begin{figure}[htbp]
\centering
\begin{tikzpicture}[active/.style={line width=1.5pt},
  ctrl/.style={line width=0.5pt, dashed}, lab/.style={font=\small}, scale=0.9]
\begin{scope}
\fill (0,0) circle (1.6pt);
\node[lab] at (0.7,0.3) {$g_v=0$};
\draw[active] (0,0) -- (1.8,0);   \node[lab] at (2.1,0) {$1$};
\draw[active] (0,0) -- (0,1.8);   \node[lab] at (0,2.05) {$1$};
\draw[ctrl] (0,0) -- (-1.5,0);    \node[lab] at (-1.75,0) {$1$};
\draw[ctrl] (0,0) -- (0,-1.5);    \node[lab] at (0,-1.75) {$1$};
\draw[ctrl] (0,0) -- (-1.1,1.1);  \node[lab] at (-1.3,1.3) {$3$};
\draw[ctrl] (0,0) -- (1.1,-1.1);  \node[lab] at (1.32,-1.32) {$\bar 3$};
\end{scope}
\begin{scope}[shift={(6.2,0)}]
\fill (0,0) circle (1.6pt);
\node[lab] at (0.7,0.3) {$g_v=0$};
\draw[active] (0,0) -- (1.8,0);   \node[lab] at (2.1,0) {$1$};
\draw[active] (0,0) -- (0,1.8);   \node[lab] at (0,2.05) {$1$};
\draw[ctrl] (0,0) -- (-1.5,0);    \node[lab] at (-1.75,0) {$1$};
\draw[ctrl] (0,0) -- (0,-1.5);    \node[lab] at (0,-1.75) {$3$};
\draw[ctrl] (0,0) -- (-1.1,1.1);  \node[lab] at (-1.3,1.3) {$1$};
\draw[ctrl] (0,0) -- (1.1,-1.1);  \node[lab] at (1.32,-1.32) {$\bar 3$};
\end{scope}
\end{tikzpicture}
\caption{Two site states in the same control class. The two active links are drawn in bold and the four control links are dashed. Both sites have the core state $s_v=(1,1;g_v=0)$. The control state is $c^1_v=(1,1,3,\bar 3)$ on the left and $c^2_v=(1,3,1,\bar 3)$ on the right.  The signed site factors agree on every transition in both directions, forward to $(3,\bar 3;g'=0)$ and $(3,\bar 3;g'=1)$ with $\sqrt 2$ and $1$, reverse from $(\bar 3,3;0)$ and $(\bar 3,3;1)$ with $\sqrt 2/3$ and $1/3$, so ADDR assigns the two environments one class label $\tilde q_v$.}
\label{fig:class-example}
\end{figure}

The observation is as follows.  The corner stages below involve lookups of the possible corner transitions, keyed on each corner's total site state (the corner source data $s_v$ and the  control link data $c_v$, see \S\ref{sec:lattice} for definitions).  There are a few hundred site states per corner at $B=6$ and about four thousand at $B=9$. Among the site states, many possess transitions that are ``essentially the same," in the following sense.  First, fix the source data $s_v$.  Each realized control pattern $c_v$ defines a collection of site factors, one value $\sigma_v w_v(s_v,c_v,t_v)$ for every allowed target $t_v$, in each direction.  Two control patterns $c^1_v$, $c^2_v$  are equivalent on the source if their site factors agree for every target in both directions, that is, $\sigma_v w_v(s_v, c^1_v, t_v)= \sigma_v w_v(s_v, c^2_v, t_v)$ for every target corner state $t_v$ and $\sigma_v w_v(t_v, c^1_v, s_v)=\sigma_v w_v(t_v, c^2_v, s_v)$ for every reverse source $t_v$.  This happens because sometimes a permutation of the control links does not change the site factors. (It would not change them at all, except for the conventional ordering of links in defining the basis of site singlets, \S\ref{sec:lattice}.) An example is shown in Fig.~\ref{fig:class-example}.

 The rotation angles looked up in the  corner stages are set by the site factors, so control patterns (on the fixed source) that are equivalent in this sense correspond to identical rotations. Therefore, we can compress the corner stage lookups by reading the equivalence class instead of the full raw site state. To do so we need a precursor circuit that computes the equivalence class, and that is the job of ADDR.

The  computation is done with one lookup per corner, mapping $(s_v,c_v)$ to its equivalence class label $\tilde q_v$ (see Table~\ref{tab:ADDR} in \S\ref{sec:cost} below). The class label is written into the KEY register. The complete action of ADDR is
\begin{equation}
\mathrm{ADDR}\,|s\rangle_{\rm SRC}|0\rangle_{\rm KEY}
=|s\rangle_{\rm SRC}\,
|\tilde q_1(s)\,\tilde q_2(s)\,\tilde q_3(s)\,\tilde q_4(s)\rangle_{\rm KEY}.
\label{eq:addr-action}
\end{equation}
Keying the corner angle tables below on $\tilde q_v$ instead of the raw control pattern reduces the number of rows by a factor of several. The four class lookups themselves contribute to the cost, and are together the largest single component of PREP at $B=9$, but it is still cheaper in the balance to use addressing rather than key the corner tables on the raw source itself. 

\subsection{Proposal stage}
\label{sec:stagea}

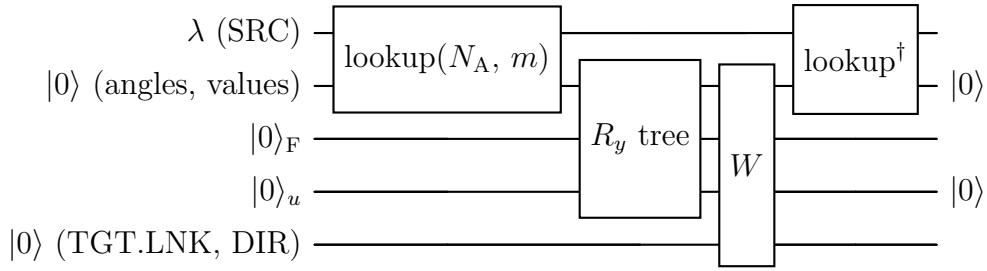
\begin{figure}[htbp]
\centering
\begin{quantikz}[column sep=0.25cm, row sep={0.7cm,between origins}]
\lstick{$\lambda$ (SRC)}
  & \gate[2]{\mathrm{lookup}(N_{\rm A},\,m)} & \qw & \qw & \gate[2]{\mathrm{lookup}^{\dagger}} & \qw \\
\lstick{$|0\rangle$ (angles, values)}
  & & \gate[3]{R_y\ \text{tree}} & \gate[4]{W} & & \rstick{$|0\rangle$}\qw \\
\lstick{$|0\rangle_{\rm F}$}
  & \qw & & & \qw & \qw \\
\lstick{$|0\rangle_{u}$}
  & \qw & & & \qw & \rstick{$|0\rangle$}\qw \\
\lstick{$|0\rangle$ (TGT.LNK, DIR)}
  & \qw & \qw & & \qw & \qw
\end{quantikz}
\caption{The proposal stage (called A for brevity in
Fig.~\ref{fig:prep-circuit}).  
See text for gate descriptions.}
\label{fig:stagea-circ}
\end{figure}

The circuit of the proposal stage is shown in Fig.~\ref{fig:stagea-circ}. It reads the active link core state $\lambda=(l_1,\dots,l_4)$  from SRC and prepares TGT.LNK and DIR in a superposition over target link cores and directions $(\lambda',d)$. The  amplitudes are given in Eq.~\eqref{eq:stagea} below. The circuit is built from a lookup of rotation angles and target $(\lambda',\mathrm{DIR})$ values, a rotation tree preparing the branch superposition on a compact index register, and a write step that transfers the indexed $(\lambda',\mathrm{DIR})$ into TGT.LNK${}\otimes{}$DIR.

Classically, we precompute the link-core transitions, the
pairs $(\lambda,\lambda')$ reached by $\Box$ or $\Box^\dagger$ for any
multiplicities and any controls.  There are about $10^3$ of these at $B=6$ and $5\cdot 10^3$ at $B=9$ (compared with tens of millions of full-core transitions, including singlet multiplicity indices, at $B=9$.)

Each proposed move is weighted high.  For corner $v$, the maximum amplitude
of an active link transition is 
\begin{equation}
\hat w_v(\lambda,\lambda') = \max_{g_v,\,c_v}\,\sum_{g_v'} w_v(\lambda,\lambda',g_v, g_v', c_v)\,.
\label{eq:what}
\end{equation}
(Recall that to get a reverse-direction transition we simply reverse the slot order of $\lambda$ and $\lambda'$ in $w$.) Here the maximum is over the corner's source multiplicity data $g_v$ and its control data $c_v$, and the sum is over the target multiplicities $g_v'$ of the allowed transitions.  Keyed on the link core state $\lambda$, the proposal stage takes 
\begin{equation}
|0\rangle_{\rm TGT.LNK}|0\rangle_{\rm DIR}|0\rangle_{\rm F}\to\sum_{(\lambda',d)}\sqrt{\frac{u_\lambda(\lambda',d)}{\hat U}}|\lambda'\rangle|d\rangle|0\rangle_{\rm F}+\sqrt{1-\frac{U_\lambda}{\hat U}}|0\rangle|0\rangle|1\rangle_{\rm F}.
\label{eq:stagea}
\end{equation}
Here 
\begin{align}
u_\lambda(\lambda',d)=\prod_{v}\hat w_v
\end{align}
 is the proposal weight of a
transition, 
\begin{align}
U_\lambda=\sum_{(\lambda',d)}u_\lambda(\lambda',d)
\end{align}
 is the total weight for a source, and 
 \begin{align}
 \hat U=\max_\lambda U_\lambda
 \end{align}
 is the  global maximum, which will return as the subnormalization in \S\ref{sec:telescope}.  The stage has the shape of Fig.~\ref{fig:stage-shape} and is shown explicitly in Fig.~\ref{fig:stagea-circ}.

The proposal stage uses classical precomputation of the reachable target link cores and directions $(\lambda',d)$ for each source link core $\lambda$,  the proposal weights $u_\lambda(\lambda',d)$, and the total $U_\lambda$. This is straightforward to do and the tables are fairly small (see Table~\ref{tab:PROP} in \S\ref{sec:cost} below.) The weights are converted classically into the  angles of the rotation tree that prepares Eq.~\eqref{eq:stagea}. Both directions, forward and reverse, are prepared by the  rotations.  The branches (basis states receiving amplitude) are the joint $(\lambda',d)$ pairs, and  the reject branch $\mathrm{F}=1$, with deficit amplitude $\sqrt{1-U_\lambda/\hat U}$.\footnote{Rejection happens at two points in PREP.  The proposal stage rejects by toggling F, as here, and each corner stage rejects by setting its own flag $\mathrm{F}_v$, \S\ref{sec:stages}.  A branch with any of these flags  set  is removed by the block encoding's final ancilla projection.}  
The angles and the targets are stored in the angle and value register slots of a QROM keyed on the four active-link irreps $\lambda$.\footnote{A row in the proposal table for a source with less than the maximum number of branches must store  zero for the unused angles in the QROM. For example, at $B=6$ the maximum number of targets $\lambda'$ for a link core state $\lambda$ is 17, but many of the other $\lambda$ have fewer than 17 targets. The values part of the row can be padded by any targets, so long as they do not collide with another physical target in the row. The simplest choice is some global reserved bitstring which is never realized; for example, a string that contains unphysical values for the link registers. (Since our truncations have six irreps per link, there are two unphysical states in the three qubit Hilbert space assigned to each link.) These rotations execute and act trivially.}

In the circuit of Fig.~\ref{fig:stagea-circ}, the first step of the rotation tree rotates F through the angle $\arccos\sqrt{U_\lambda/\hat U}$.  The later stages are controlled on $\mathrm{F}=0$, so a rejected branch passes through unmodified.   The write step $W$ transfers the superposition prepared on the branch number $u$ to the corresponding states of TGT.LNK representing the actual target link values, which are read from the value register.  It is the general write circuit of \S\ref{sec:data-quantum} (Fig.~\ref{fig:write-w}).

\subsection{Corner stages}
\label{sec:stages}

\begin{figure}[htbp]
\centering
\resizebox{\linewidth}{!}{%
\begin{quantikz}[column sep=0.3cm, row sep={0.8cm,between origins}]
\lstick{$(s_v,\tilde q_v,\mathrm{DIR},\,l_i',l_j')$}
  & \gate[2]{\mathrm{lookup}(N_v,\,m)} & \gate[2]{\delta^{\rm row}_v} & \qw & \qw & \qw & \qw & \qw & \gate[2]{\delta^{\rm row}_v} & \gate[2]{\mathrm{lookup}^{\dagger}} & \qw \\
\lstick{$|0\rangle$ (key image, angles)}
  & & & \octrl{4} & \qw & \qw & \gate[3]{R_y\ \text{tree}} & \octrl{4} & & & \rstick{$|0\rangle$}\qw \\
\lstick{$|0\rangle_{\mathrm{F}_v}$}
  & \qw & \qw & \qw & \targ{} & \targ{} & & \qw & \qw & \qw & \qw \\
\lstick{$|0\rangle_{g_v'}$ (TGT.G)}
  & \qw & \qw & \qw & \qw & \qw & & \qw & \qw & \qw & \qw \\
\lstick{$\mathrm{F}$}
  & \qw & \qw & \octrl{1} & \octrl{-2} & \qw & \qw & \octrl{1} & \qw & \qw & \qw \\
\lstick{$|0\rangle_{\mathrm{c}_v}$}
  & \qw & \qw & \targ{} & \qw & \ctrl{-3} & \ctrl{-4} & \targ{} & \qw & \qw & \rstick{$|0\rangle$}\qw
\end{quantikz}%
}
\caption{Corner stage $B_v$ writes the multiplicity $g_v'$ or toggles the reject flag
$\mathrm{F}_v$, with amplitudes set by $\sqrt{w_v/\hat w_v}$, keyed on the corner source data $s_v$
(read from SRC), the control class $\tilde q_v$ (read from KEY), DIR, and the corner's two target
links (read from TGT.LNK).  The lookup returns the row's key and the angles, and the input key is XORed onto the output key, computing the key difference $\delta^{\rm row}_v$.  The proposal flag F and $\delta^{\rm row}_v$ are ANDed together into the accept bit
$\mathrm{c}_v$, and two CX gates set $\mathrm{F}_v=\bar{\mathrm{F}}\oplus{\mathrm{c}}_v$ ($=\bar{\mathrm{F}}\wedge\bar{\mathrm{c}}_v$ on the realizable inputs). Rotations are controlled on $c_v$ and $\mathrm{F}_v$ so that keys absent from the corner table do not generate any rotations, but are flagged. }
\label{fig:corner-circ}
\end{figure}
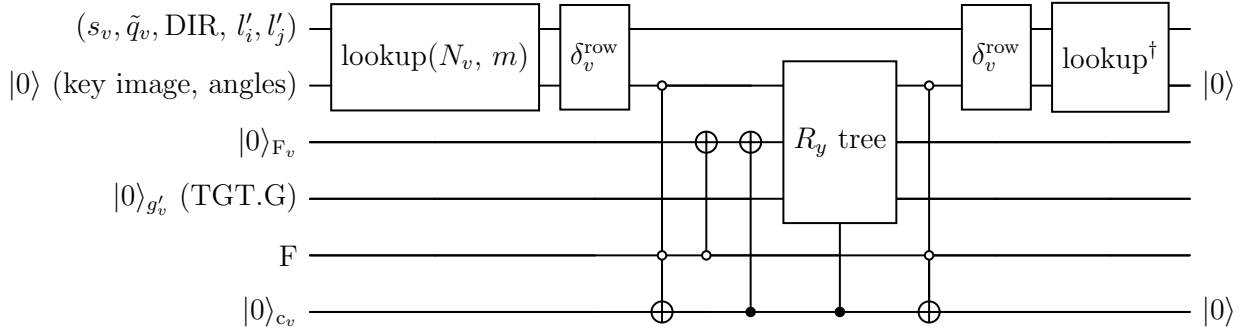

The job of the corner stages is to correct the proposed amplitudes based on each corner's actual site state. The general form of the circuit is shown in Fig.~\ref{fig:corner-circ}. 
 Keyed on the source data $s_v$ (read  from SRC), the control class $\tilde q_v$ (read from KEY), the corner's two target links (read from TGT.LNK), and DIR, and conditioned on $\mathrm{F}=0$, the corner stage rotates the corner's target singlet multiplicity register (the corner's slot in TGT.G) and its reject flag qubit $\mathrm{F}_v$, as:
\begin{equation}
|0\rangle_{\mathrm F}|0\rangle_{\mathrm F_v}|0\rangle_{g_v'}\to|0\rangle_{\mathrm F}\Bigl[\sum_{g_v'}\sigma_v(g_v')\sqrt{\tfrac{w_v}{\hat w_v}}|0\rangle_{\mathrm F_v}|g_v'\rangle+\sqrt{1-\tfrac{\tilde w_v}{\hat w_v}}|1\rangle_{\mathrm F_v}|0\rangle_{g_v'}\Bigr]\qquad
\tilde w_v=\sum_{g_v'} w_v
\label{eq:corner-stage}
\end{equation}
Here the sum runs over the target multiplicities consistent with the corner's environment.  The flag  $\mathrm{F}_v$ plays the role for corner $v$ that F plays for the proposal, and the stage acts as the identity on the F $=$ 1 sector.  As in the proposal tree, the corner tree's first rotation acts on $\mathrm{F}_v$, and subsequent multiplicity index rotations are conditioned on $\mathrm{F}_v=0$. Each accepted amplitude is the square root of the ratio of the true weight to the optimistic proposed one. More precisely, (\ref{eq:corner-stage}) is the action of $\mathrm{PREP}$, whereas for $\mathrm{PREP}'$,  the sign  $\sigma_v$ is dropped. As a consistency check, we can see that Eq.~\eqref{eq:what} implies $\tilde w_v\le\hat w_v$. 

A proposed target link core and direction $(\lambda',\mathrm{DIR})$ with no consistent multiplicity at this corner has $\tilde w_v=0$, and the register must have its flag flipped to $\mathrm{F}_v=1$ on this branch. Like $\mathrm{F}=1$, branches of the wavefunction with $\mathrm{F}_v=1$ are marked failed and  removed by the block encoding's ancilla projection. One method to implement the rejection is to include explicit zero-lines in the tables for invalid transitions, but this is expensive, since there are  many of them. On the other hand, if we leave invalid transitions out of the table, the proposal stage will propose them anyways, and the sparse lookup described in Appendix~\ref{app:sparse-qrom} will miss-key, returning some corrupted result. We leave invalid transitions out of the tables, and insert an extra check to make sure that the key passed to the sparse lookup matches the row it returns, flipping to the reject branch if  not. 

The classical precomputation needed to construct the corner stages is  similar to the proposal stage.  For each  key (source data $s_v$, class $\tilde q_v$, a target link pair, and a direction), the classical corner tables enumerate the allowed target site multiplicity indices $g_v'$, their site factors $w_v$, the marginal $\tilde w_v$, and the ceiling $\hat w_v$ of Eq.~\eqref{eq:what} (see Table~\ref{tab:CORN} in \S\ref{sec:cost} below.) The lookups return the  rotation tree angles, at $b_{\rm rot}$ bits per angle, and a copy of the key for the validation described above. There is one table per corner, with $k_f$ rows in the forward direction and $k_r$ in the reverse, and the key includes a direction bit.

After the lookup is called we perform the key check. One  CX per bit XORs the input key onto the output key register, which leaves that register in the all-zeros state iff the keys agree. An open-controlled Toffoli tree ANDs together the negated bits of the output key register and $\neg F$ into a bit $c_v$, and two CX gates toggle the corner flag $\mathrm{F}_v$. After the check, the possible states of the accept bit and corner flag are:
\begin{equation}
(\mathrm{c}_v,\,\mathrm{F}_v)=
\begin{cases}
(1,\,0), & \text{if}\ \mathrm{F}=0 \ \text{and}\ \delta^{\rm row}_v=0,\\
(0,\,1), & \text{if}\ \mathrm{F}=0 \ \text{and}\ \delta^{\rm row}_v\neq0,\\
(0,\,0), & \text{if}\ \mathrm{F}=1.
\end{cases}
\label{eq:corner-check}
\end{equation}
After the key check we perform the programmed rotation tree. The first rotation puts the correct deficit weight into the $|1\rangle_{\mathrm{F}_v}$, controlled on $c_v$, so it only acts on the $|0\rangle_{F}$ branch of the proposal. All subsequent rotations are of the multiplicity register, and are controlled on both $c_v=1$ and ${\mathrm{F}_v}=0$, so that rotations only happen on the $\mathrm{F}_v=0$, F $=$ 0 branch and correspond to valid transitions. Finally $c_v$ is uncomputed.   The net action of the stage is
\begin{equation}
B_v:\quad
\begin{cases}
\text{Eq.~\eqref{eq:corner-stage}}, & \delta^{\rm row}_v=0,\\[2pt]
|0\rangle_{\mathrm F}\,|0\rangle_{\mathrm F_v}|0\rangle_{g_v'}\;\to\;
|0\rangle_{\mathrm F}\,|1\rangle_{\mathrm F_v}|0\rangle_{g_v'}, & \delta^{\rm row}_v\neq0,\\[2pt]
|1\rangle_{\mathrm F}|\psi\rangle\;\to\;|1\rangle_{\mathrm F}|\psi\rangle. &
\end{cases}
\label{eq:corner-cases}
\end{equation}
The middle line is the $\tilde w_v\to0$ limit of Eq.~\eqref{eq:corner-stage}.\footnote{Some of the flagging we are doing here is overkill for the purpose of block encoding. It is not strictly necessary, for example, that the oracle acts as the identity operator on the F $=$ 1 branch of the proposal flag, since all this weight will be projected out by the mechanism of \S\ref{sec:correctness}. Nonetheless we implement it to keep the bookkeeping more straightforward.}

Each of the five flags, F and the four $\mathrm{F}_v$, is associated with one qubit of the tag register ${\rm TAG}$  in Fig.~\ref{fig:prep-circuit}.  After the rotation tree of each stage, we write ${\rm TAG}=\mathrm{F}\cdot\mathrm{DIR}$ and ${\rm TAG}_v=\mathrm{F}_v\cdot\mathrm{DIR}$  with a Toffoli. (This is not shown explicitly in Figs.~\ref{fig:stagea-circ} or \ref{fig:corner-circ}.)  On a reject branch, the tag records the direction at the moment of rejection. On  other branches the Toffoli does nothing.  The tags are critical  because the oracle flips the DIR bit between the two prepares. Recording the rejection direction ensures that the reject branches of the bra and ket have zero overlap with each other, as discussed further in \S\ref{sec:correctness}. Flags and tags sit in the ancilla register $a$  whose projection onto $|0\rangle_a$ removes rejected branches.


The four stages are mutually independent given $(\lambda',\mathrm{DIR})$, since each reads only SRC, KEY, TGT.LNK, and DIR and writes only its own TGT.G slot, so they can be run in any order. The unprepare $\mathrm{PREP}'^{\dagger}$ has the same  layout as the prepare but stores the unsigned corner angles. It is keyed on SRC, which holds the neighbor after the swap.  Its role is discussed further in \S\ref{sec:finishing} below.

\subsection{Combining stages of PREP}
\label{sec:telescope}

On a branch of the wavefunction with all flags unset, the stages of PREP compose to give the basis state $|\lambda'\rangle_{\rm TGT.LNK}|g_1'\cdots g_4'\rangle_{\rm TGT.G}|d\rangle_{DIR}$ amplitude
\begin{equation}
\sqrt{\frac{\prod_v \hat w_v}{\hat U}}
\prod_{v=1}^{4}\sigma_v\sqrt{\frac{w_v}{\hat w_v}}
=\sigma_{st}\sqrt{\frac{w_1w_2w_3w_4}{\hat U}}
=\sigma_{st}\sqrt{\frac{|\Box_{st}|}{\alpha_p}}\,,
\qquad \alpha_p=\hat U,
\label{eq:telescope}
\end{equation}
using the  factorization $|\Box_{st}|=w_1w_2w_3w_4$ of \S\ref{sec:lattice}.  Thus the subnormalization is $\alpha_p=\hat U$. Analyzing the site factors, we find $\hat U=8.6$ at $B=6$ and $\hat U=56.2$ at $B=9$. 

\subsection{Other planes}
\label{sec:otherplanes}

The construction so far describes one plaquette orientation, but in three dimensions, there are three orientations. The orientations are related by cubic rotations, which preserve the spectrum, so $\|H_p\|$ is common to all three. Unfortunately, however, the row-$1$-norm and the ceilings built from the site factors are not spectral quantities, and the canonical  $F$-ordering in the basis of singlet wavefunctions (CGCs) is not invariant under rotations. For this reason there are variations  in the plane-to-plane site factors. Measured per plane, the proposal maximum $\hat U$ is plane-independent at $B=6$, while at $B=9$ there are slight variations. The construction uses  $\alpha_p=\max_{\rm plane}\hat U$ for every plaquette.

The classical tables are orientation-dependent for the same reason, and routing already supplies the plaquette index $p$, so extracting the plane and including it in the keys of the lookups is straightforward.  The tables differ only in how much of their content merges across the planes.  For the class lookups and the proposal stage,  one lookup serves all three planes with height given by the sum of the three per-plane heights.   The corner stages need no explicit plane bits in the keys because their key is the class label, which ADDR already merges over the planes. Corner rows that  appear verbatim in different planes are only stored once. 

\section{Completing the magnetic block encoding}
\label{sec:finishing}

\subsection{SWAP}
\label{sec:swap}

The plaquette operator is completely off-diagonal in the electric basis. After PREP, the target neighbor core states $t$ have been written into TGT.  To implement the off-diagonal structure we must swap TGT with the core fields of the source state $s$ in SRC. X(DIR) then flips the direction (a transition that is forward for the source is reverse for the neighbor), and PREP$'^{\dagger}$, keyed on the neighbor now sitting in SRC, uncomputes TGT. SWAP and X(DIR) involve purely Clifford gates.

\subsection{\texorpdfstring{${\rm\bf PREP}^{\prime\dagger}$}{Prepprimedagger}}
\label{sec:sign}

As discussed above, PREP${}^{\prime}$ is identical to PREP except that its stored corner angles prepare states with amplitude $\sqrt{\frac{w_v}{\hat w_v}}$ instead of $\sigma_v \sqrt{\frac{w_v}{\hat w_v}}$.  The class lookups, the proposal table, and all table heights are identical in both, so the two prepares will have the same  $T$-gate cost, estimated below.

\subsection{Full magnetic oracle}
\label{sec:correctness}
Let us track the state through the elements of the $\Box+\Box^\dagger$ oracle. We have:
\begin{equation}
\begin{aligned}
&|s\rangle_{\rm SRC}|0\rangle_{\rm DIR}|0\rangle_{\rm TGT}|0\rangle_{\rm F}\\
&\xrightarrow{\ \mathrm{PREP}\ }
\textstyle\sum_{t}\sigma_{st}\sqrt{\tfrac{|h_{st}|}{\alpha_p}}|s\rangle_{\rm SRC}|d_t\rangle_{\rm DIR}|\mathrm{core}(t)\rangle_{\rm TGT}|0\rangle_{\rm F}|0\rangle_{{\mathrm F}_v}+(\text{reject})\\
&\xrightarrow{\ \mathrm{SWAP}\ }
\textstyle\sum_{t}\sigma_{st}\sqrt{\tfrac{|h_{st}|}{\alpha_p}}|t\rangle_{\rm SRC}|d_t\rangle_{\rm DIR}|\mathrm{core}(s)\rangle_{\rm TGT}|0\rangle_{\rm F}|0\rangle_{{\mathrm F}_v}+(\text{reject})\\
&\xrightarrow{\ \mathrm{X(DIR)}\ }
\textstyle\sum_{t}\sigma_{st}\sqrt{\tfrac{|h_{st}|}{\alpha_p}}|t\rangle_{\rm SRC}|\bar d_t\rangle_{\rm DIR}|\mathrm{core}(s)\rangle_{\rm TGT}|0\rangle_{\rm F}|0\rangle_{{\mathrm F}_v}+(\text{reject}),
\end{aligned}
\label{eq:prep-trace}
\end{equation}
where the sum runs over neighbors $t$ of $s$ (forward neighbors with $d_t=0$, reverse neighbors with $d_t=1$) and $\bar d_t$ is the complementary bit. Finally the action of $\mathrm{PREP}'^{\dagger}$ on $\langle t';0|$ is:
\begin{align}
&|t'\rangle_{\rm SRC}|0\rangle_{\rm DIR}|0\rangle_{\rm TGT}|0\rangle_{\rm F}|0\rangle_{{\mathrm F}_v}\nonumber\\
&\xrightarrow{\ \mathrm{PREP'}\ }
\textstyle\sum_{s'}\sqrt{\tfrac{|h_{t's'}|}{\alpha_p}}|t'\rangle_{\rm SRC}|d_{s'}\rangle_{\rm DIR}|\mathrm{core}(s')\rangle_{\rm TGT}|0\rangle_{\rm F}|0\rangle_{{\mathrm F}_v}+(\text{reject}).
\label{eq:prepdagger}
\end{align}
Taking the overlap of (\ref{eq:prep-trace}) and (\ref{eq:prepdagger}) picks out $s=s'$ and $t=t'$, and by Hermiticity $|h_{ts}|=|h_{st}|$ the  block entry collapses to
\begin{equation}
\langle t,0|\,U_p\,|s,0\rangle
=\sigma_{st}\,\sqrt{\tfrac{|h_{st}|}{\alpha_p}}\,\sqrt{\tfrac{|h_{ts}|}{\alpha_p}}
=\frac{h_{st}}{\alpha_p}\,.
\label{eq:prep-multiply}
\end{equation}
The only check we have to make is that the two reject sectors in (\ref{eq:prep-trace}) and (\ref{eq:prepdagger}) have zero overlap. The overlap vanishes because the two sides always disagree in their direction data.  A corner reject in Eq.~\eqref{eq:prep-trace} was tagged with the direction it was rejected under, and then had DIR flipped by the X, so it appears in the overlap with ${\rm TAG}_v\neq\mathrm{DIR}$. The corner rejects in Eq.~\eqref{eq:prepdagger} have ${\rm TAG}_v=\mathrm{DIR}$. The same properties hold for the proposal flag and tag.  Thus the bra and ket reject branches  disagree either in the direction or in a tag, and the overlaps vanish.

\section{Costs}
\label{sec:cost}

In this section we tally the $T$ gate costs of one call to the single-plaquette oracle that block encodes $\Box_p+\Box_p^\dagger$ in SU(3) gauge theory, and the logical qubits needed to run the algorithm.  In a fault-tolerant architecture often the Clifford gates are comparatively cheap, while each $T$ gate has to be produced by magic state distillation or similar procedures, and the latter dominate the  cost.  We use the rate ${\rm Toffoli} =4\,T$.
Where possible, Toffolis computed into ancillas can be uncomputed by  measurement-based methods, without additional $T$ gates~\cite{gidney2018halving}. 

Some raw data is collected in Table~\ref{tab:raw}.
\begin{table}[htbp]
\centering\small
\begin{tabular}{l|ccc}
\hline
Truncation & site states & control patterns & site factors \\
\hline
B=6 & 424  & 109 & 473 \\
B=9 & $\approx 4250$ & 475 & $\approx10^4$ \\
\hline
\end{tabular}
\caption{Properties of the truncations. Counts correspond to a single corner and plane. $\approx$ denotes that the numbers vary slightly with plane and corner, a consequence of the basis construction of the site singlets.}
\label{tab:raw}
\end{table}
From the site states and factors we can construct the classical tables used throughout  the single-plaquette oracle. The rows and output bits of the ADDR tables are shown in Table~\ref{tab:ADDR}; similarly for the proposal table in Table~\ref{tab:PROP}, and for the corner tables in Table~\ref{tab:CORN}.

\begin{table}[htbp]
\centering\small
\begin{tabular}{l|ccc}
\hline
Truncation & ADDR  rows & max ctrl classes & label bits   \\
\hline
B=6 & 1518  & 94 & 7 \\
B=9 & 13638 & 1943 & 11 \\
\hline
\end{tabular}
\caption{ADDR table rows and output bits, merged over the three planes and maximum over corners. Keys $=$ site states $+$ plane, output $=$ control class; 4 tables total for 4 corners.}
\label{tab:ADDR}
\end{table}

\begin{table}[htbp]
\centering\small
\begin{tabular}{l|ccc}
\hline
Truncation & Proposal rows & max target states/row & output bits   \\
\hline
B=6 & 1212  & 17 &  17$\times(b_{rot}+13)$\\
B=9 & 3174 & 32 & 32$\times(b_{rot}+13)$ \\
\hline
\end{tabular}
\caption{Proposal table rows and output bits, merged over the three planes. $b_{rot}$ is the angle precision in bits, and $13=3$ qubits/link $\times$ 4 links $+$ 1 direction bit. Keys $=$ link core states $+$ plane, output $=$ angles for preparing a superposition over the link core's neighbors $+$ the bitstrings of those neighbors $+$ the direction bits; 1 table.}
\label{tab:PROP}
\end{table}

\begin{table}[htbp!]
\centering\small
\begin{tabular}{l|ccc}
\hline
Truncation & Corner rows & max target states/row & max output bits.    \\
\hline
B=6 & 244  & 2 &  2$\times b_{rot}$\\
B=9 & 6972 & 6 & 6$\times b_{rot} $ \\
\hline
\end{tabular}
\caption{Corner table rows and output bits, counts maximized over the four corners. $b_{rot}$ is the angle precision in bits.  Keys $=$ corner core source state $+$ corner class $+$ target link pair $+$ direction, output $=$ angles for preparing a superposition over the site singlet multiplicity index states; 4 tables total for 4 corners.}
\label{tab:CORN}
\end{table}

Now we can begin to estimate the logical qubit and gate costs associated with the oracle. Each QROM and its uncompute  appearing in ADDR, the proposal stage, and the corner stages  contributes $4(N_{\rm stage}-1)\, T$. These costs are tallied approximately in the first three rows of Table~\ref{tab:Tcost} (counts are only approximately $8m_{tbl}(N_{\rm stage}-1)$, where $m_{tbl}$ is the multiplicity of the tables and the various $N_{\rm stage}$ taken from Tables~\ref{tab:ADDR},~\ref{tab:PROP},~\ref{tab:CORN}, due to variations among the corners.) The programmed rotation  trees in the proposal and corner stages contribute roughly $k\times (b_{rot}+\log_2(k)-1)$ Toffolis, where $k$ is the number of angles in the tree, $b_{rot}$ is the angle precision, $k b_{rot}$ is the contribution from the adder, and $k(\log_2(k)-1)$ well-approximates the number of controls in the tree. The dominant cost of the write part of the proposal stage is the $k\times 13$ Toffolis in the ``write j" gate (controlled-copying $13=3$ qubits/link $\times$ 4 links $+$ 1 direction bit into the target register) and the same number to uncompute the flag qubit. The corner stage key check is one Toffoli for each compared bit of the key, per corner.

\begin{table}[htbp]
\centering\small
\begin{tabular}{l|ccc}
\hline
 & B=6 & B=9     \\
\hline
ADDR QROM & 48500  & 436000\\
Proposal QROM & 9700 & 25400 \\
Corner QROM & 7400 & 157000 \\
Proposal Rotation & 216 + 68 $b_{rot}$ &516 + 128 $b_{rot}$ \\
Corner Rotation &32 + 32 $b_{rot}$ & 392 + 88 $b_{rot}$\\
Corner key check & 220 & $\sim 10^3$\\
Proposal Write & 1768 & 3328 \\
\hline
PREP & $6\times 10^4$ & $6\times 10^5$ \\
$U_p$ & $1.2\times 10^5$ & $1.2\times 10^6$ \\
\hline
\end{tabular}
\caption{Estimated $T$-counts for the different elements of the single plaquette oracle. Numbers for the QROM lookups have been rounded. $b_{rot}$ is the precision with which rotation tree angles are stored. Its contribution is sufficiently small that it does not enter the total estimate for PREP. The $T$ count of $U_p$ is that of PREP + PREP$^{'\dagger}$, which is twice PREP.}
\label{tab:Tcost}
\end{table}

The logical qubit requirements are dominated by the lattice state register and the angle and value registers holding the output of the proposal lookup. Each lattice link needs 3 qubits to cover the six irrep states, and 1 (3) qubits to encode the site singlet multiplicity index states at $B=6$ ($B=9$). The counts are tallied in Table~\ref{tab:qcounts}.

\begin{table}[htbp]
\centering\small
\begin{tabular}{l|ccc}
\hline
 & B=6 & B=9     \\
\hline
Lattice state & $10L^3$  & $12L^3$\\
Proposal angle anc. & 17 $b_{rot}$ & 32 $b_{rot}$ \\
Proposal value anc. & $13\times17$  & $13\times32$  \\
\hline
\end{tabular}
\caption{Main  logical qubit allocations. Other ancilla registers are much smaller, so the sum of these three rows gives the estimated requirements.}
\label{tab:qcounts}
\end{table}

We see that the qubit requirements are similar between the truncations, but the $T$ counts grow substantially, driven by the lookup tables. We comment on directions for optimization in the conclusions.

\section{Explicit circuits}

For validation purposes, the oracle assembled in the preceding sections was implemented and compiled in Guppy~\cite{Koch:2025qnc}.  A single-plane build at $B=6$ with $b_{\rm rot}=38$ allocates $1092$ qubits, of which $147$ are persistent and the remainder is the transient workspace  at the widest point of the circuit:
\begin{center}
\begin{tabular}{lr}
\hline
persistent & \\
\quad SRC links and multiplicities (4 active + 16 control, 3 qubits each; 4 mult) & 64 \\
\quad TGT links and multiplicities ($12+1$ DIR; 4) & 17 \\
\quad flags $F$, $F_v$, ${\rm TAG}$ & 10 \\
\quad class slots & 11 \\
\quad phase gradient and guard & 39 \\
\quad branch index and write flag & 6 \\
peak transient (proposal stage) & \\
\quad lookup data register, $k(b_{\rm rot}+v_b) = 17\times 51$ & 867 \\
\quad programmed rotation working qubits & 41 \\
\quad phase-gradient adder carries, $b_{\rm rot}-1$ & 37 \\
\hline
total & 1092 \\
\hline
\end{tabular}
\end{center}
\noindent The workspace data register dominates and can be reused on successive plaquettes. 

The gate counts agree with the findings of Sec.~\ref{sec:cost}. Despite having a large qubit count, the compiled circuit can also be run in the following sense. Given a computational basis input, we follow its path through the lookups and create new branches at the programmed rotations. In this way we accumulate a sparse table of basis states with nonzero amplitude. Every tabulated transition was checked in this way, fifteen control sectors per plane and $1698$ transitions over the three planes, and the amplitudes $\langle 0, t|\,U_p\,|0, s\rangle$ reproduce $H_{st}/\hat{U}$. 

Register states outside the physical basis are found to carry no amplitude in the accept subspace. Two classes were prepared, a control pattern carrying the reserved label of Sec.~\ref{sec:address} and a link core outside the proposal domain of Sec.~\ref{sec:stagea}. In both cases the projection argument of Sec.~\ref{sec:correctness} is realized on the compiled circuit.

The compiled circuits, data tables, and a standalone verifier implementing the sparse simulation described above are available from Ref.~\cite{magneticoracle}.

\section{Usage}
\label{sec:qsvt-vs-trotter}

One use of the block encoding oracle is in a QSVT implementation of the time evolution operator $e^{-iHt}$, cf. Figs.~\ref{fig:qsvt} and~\ref{fig:projphase}. 
\begin{figure}[htbp]
\centering
\begin{quantikz}[column sep=0.35cm, row sep={0.9cm,between origins}]
\lstick{$|0\rangle_{a}$}
  & \gate{e^{i\phi_0 (2\Pi-I)}} & \gate[2]{U_H} & \gate{e^{i\phi_1 (2\Pi-I)}} & \gate[2]{U_H^\dagger}
  & \gate{e^{i\phi_2 (2\Pi-I)}} & \ \cdots\  & \gate[2]{U_H} & \gate{e^{i\phi_d (2\Pi-I)}} & \qw \\
\lstick{$|\psi\rangle$}
  & \qw & & \qw & & \qw & \ \cdots\  & & \qw & \qw
\end{quantikz}
\caption{High-level QSVT circuit.} 
\label{fig:qsvt}
\end{figure}
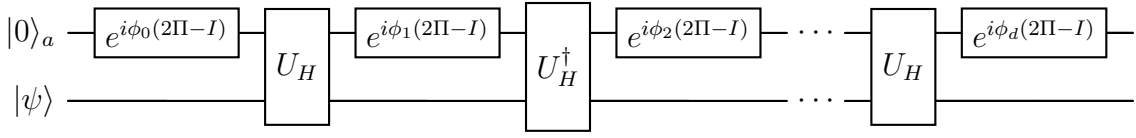
\begin{figure}[htbp]
\centering
\begin{quantikz}[column sep=0.6cm, row sep={0.95cm,between origins}]
\lstick{$|0\rangle_a$} & \qw & \octrl{1} & \qw & \octrl{1} & \qw \\
\lstick{$|0\rangle_{\rm flag}$} & \qw & \targ{} & \gate{e^{\,i\phi_k(2|1\rangle\langle1|-I)}} & \targ{} & \qw
\end{quantikz}
\caption{The projector-controlled phase $e^{i\phi_k(2\Pi-I)}.$}%
\label{fig:projphase}
\end{figure}
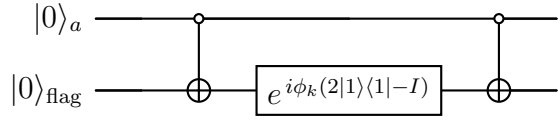
Given $U_H$, QSVT implements a degree-$d$ polynomial $p(H/\beta)\approx
e^{-iHt}$.  Conceptually, $U_H^\dagger$ takes the ``A qubit" subspaces, spanned by $|0\rangle_a|\lambda_i\rangle$  and $U_H |0\rangle_a|\lambda_i\rangle$, to the ``B qubit" subspaces, spanned by $|0\rangle_a|\lambda_i\rangle$ and $ U_H^\dagger |0\rangle_a|\lambda_i\rangle$, and $U_H$ takes the B qubit subspaces to the A qubit subspaces; here $|\lambda_i\rangle$ denote the eigenvectors of $H$.  The phase rotations act like $R_Z$ on each A and B qubit, moving around on each Bloch sphere. By interleaving rotations and flips we may effect an arbitrary polynomial $p(H/\beta)$ in the upper left block.
 
The phase rotations are realized by one extra flag qubit (not shown above) and two multi-controlled gates, cf. Fig.~\ref{fig:projphase}. 

For large subnormalization $\beta$, the $T$-cost may be estimated as~\cite{low2019qubitization,gilyen2019qsvt}
\begin{align}
T_{\rm QSVT}&\approx 2\beta t\,[T(U_p)+T(U_p^\dagger)+T_{\rm route}(L)+T(U_E)].
\label{eq:qsvtcost}
\end{align}
The contribution from $T(U_p)$ was estimated in the final line of Table~\ref{tab:Tcost} above. A full lattice has $3L^3$ plaquettes which must be routed to the single-plaquette oracle, so $T_{\rm route}(L)\sim 10^2 L^3$. (The $10^2$ is a crude estimate of the cost of shift-swapping one plaquette to the oracle.) The prefactor $\sim\beta t$ estimates the required degree of the polynomial approximation to $e^{-iHt}$ implemented by QSVT, and since $\beta$ is the subnormalization of the whole Hamiltonian it is at least $3L^3 \alpha_p$. Thus for large $L$ we have
\begin{align}
T_{QSVT}\sim 10^2 \alpha_p L^6 t 
\end{align}
entirely from routing and the polynomial degree. At the truncations we considered routing dominates the plaquette oracle above $L=20$ or so. Further improvements to the oracle above do not mitigate this term. 

Another use, with different scaling, is  a hybrid Trotter-QSVT algorithm. This is an ordinary Trotterization in which the exponentials are built using QSVT with the same oracle.  Partition the lattice into $M$ cubic chunks of side $\ell$, Trotterize over the electric term and the chunks, and implement each chunk exponential, 
\begin{equation}
e^{-i\,\delta t\,H_c},\qquad
H_c=\frac{1}{g^2}\sum_{p\in c}\big(\Box_p+\Box_p^\dagger\big),
\label{eq:chunk-exp}
\end{equation}
with the  QSVT circuit of Fig.~\ref{fig:qsvt}. The block
encoding of each chunk has subnormalization $\beta_c=3\ell^3\alpha_p$ and  routing
confined to the chunk (plus the control
links of the boundary plaquettes that are normal to the boundary). 

In an extreme case, we place each plaquette into its own chunk. In this case the routing is deleted and we  simply implement $e^{-i\,\delta t\,H_p}$ with QSVT. To  accuracy $\epsilon'$, the required number of queries to the oracle is~\cite{gilyen2019qsvt} (the second term was omitted above, but must be retained here since it will be enhanced in~(\ref{eq:trotter-oracle-total}))
\begin{equation}
Q_p \sim 2\alpha_p\,\delta t + 4\sqrt{\alpha_p\,\delta t\,\log(1/\epsilon')}\,.
\label{eq:trotter-oracle-degree}
\end{equation}
This is a factor of $1/(3L^3)$ smaller than the global query count.
Longer time evolution is then implemented by the standard second-order product formula together with the electric
term $V_E=e^{-i\,\delta t\,H_E}$. The step count $r$ is related to 
 the Trotter error. The worst case is $r=O\big(\sqrt{N_{\rm plaq}\,W\,t^3/\epsilon}\big)$ at second order, where $W$ is a commutator norm~\cite{childs2021trotter}, but the reality might be much better. If the total polynomial-approximation error across the evolution is $\epsilon$, then the budget per  exponential is $\epsilon'=\epsilon/(rN_{plaq})$, and the cost of the hybrid algorithm is
\begin{align}
T_{\rm Trot}&\approx 2\alpha_B\,t\,\left[T(U_{p})+T(U_p^\dagger)\right]+
4\sqrt{\alpha_p\,\delta t\,\log(1/\epsilon')}\,r\,N_{plaq}\,\left[T(U_{p})+T(U_{p}^\dagger)\right]+r\,T(V_E) .
\label{eq:trotter-oracle-total}
\end{align}
 The first term of  Eq.~\eqref{eq:trotter-oracle-total} uses $r\,N_{\rm plaq}\,\alpha_p\,\delta t=\alpha_B\,t$, and it is identical to the oracle term in the ordinary QSVT cost.  The differences from QSVT are in the electric encoding, the routing specific to QSVT, and the polynomial error on each of the $r\,N_{plaq}$ QSVT-approximated exponentials in the Trotterization. In the asymptotic large-$L$ limit the scaling of the Trotter approach is better, $L^{4.5}$ (taking the worst-case $r$) versus $L^6$, but the prefactor is different, and depends both on the details of the Trotter error and the choice of the chunk size $M$, if we restore that freedom. 
 
One useful feature of the Trotter approach is increased parallelization, if qubits are cheap.  With a large enough working register of qubits per unit plaquette volume, the depth per Trotter step is independent of $L$ and the total depth scales as $r\sim L^{1.5}t^{3/2}$, compared with a QSVT depth that scales roughly as $L^6 t\log L$ (number of queries, routing depth). 

A proper comparison requires careful estimation of the Trotter error, explicit treatment of the electric Hamiltonian, proper accounting of chunk-boundary control links,  further opportunities for optimizing $T(U_p)$, and other investigations which  which we defer to future work. With these in hand, however, it should be possible to determine which algorithm/choice of $M$ optimizes gate counts and depth at the $L$ values of practical interest.

\section{Discussion}
\label{sec:conclusions}

We conclude with brief comments on directions for future work. 

\begin{itemize}
\item Reaching the continuum at $g=1$ may require raising the per-vertex electric energy truncation used here beyond $B=9$ (or implementing a different truncation). The classical precomputation can be extended accordingly; for example, corner tables grow by a factor of about $300$ in rows at $B=16$. To lower the oracle cost in this case, it will be advantageous to explore optimization opportunities we have ignored. Key compression before lookup to take advantage of better algorithms for densely-keyed lookups is one avenue. (See, for  example,~\cite{li2026sparseqrom}, which uses hashing to implement a sparse QROM with $T$ count of order roughly 
$\sqrt{{\rm rows}\times{\rm bits}}$.)  Another is replacing parts of ADDR with in-register arithmetic, recognizing that to a significant degree the control classes are governed by subgroups of permutation symmetries.
\item The Trotter-QSVT hybrid algorithm described in \S\ref{sec:qsvt-vs-trotter} needs further exploration. Whether its  tradeoffs win at accessible volumes depends on realistic Trotter error analysis and further optimization of the single plaquette oracle. 
\item  We can readily extend the analysis to include the matter sector of full QCD.  The gauge-matter interaction also
factorizes into site factors in the irrep basis~\cite{hidalgo2026lqcd}, so the
lookup-based construction developed here is directly applicable. 
\end{itemize}

With these ingredients in hand,
a complete and precise estimate for the resource cost of   fault-tolerant simulations of QCD at physical truncations  should be possible.

~\\
{\emph{Note added:} Shortly after this work first appeared, Davoudi and Stryker~\cite{Davoudi:2026pcu} gave an improved
product-formula treatment of the same magnetic term in the electric basis of~\cite{byrnes2006simulating,kan2021lqcd}.  They reduce the number of subevolutions a Trotter step must
sequence by a factor exponential in $N_c^2$. Unlike the classical precomputation and lookup table approach advocated here, all transition amplitudes are assumed to computed by in-register CGC arithmetic along the lines of~\cite{kan2021lqcd}.  A full comparison of end-to-end costs for this basis and implementation, versus the gauge-invariant basis used here, remains well motivated work for the future.}

\section*{Acknowledgments}

I thank Luis Hidalgo for useful conversations. This work was supported by the U.S. Department of Energy, Office of Science, Office of High Energy Physics Quantum Information Science Enabled Discovery (QuantISED) program. This work was performed in part using LLMs. All results are the responsibility of the author. 

\appendix

\section{Lookups by sparse unary iteration}
\label{app:sparse-qrom}

The lookups in PREP encode classical tables where the rows are indexed by a physical key, and the keys that actually occur are a sparse subset of a much larger space of possible bit patterns.  Sparse unary iteration allows efficient reading of the table. It works as follows. 

First, we build a sequence of logical operators that uniquely identifies each key, under the promise that only the keys of the table will ever be passed to the QROM.\footnote{In cases where it is possible that a QROM will be a passed an unrecognized key, an extra check is needed to deal with it. This happens in the corner stage of the main text. In that case the output contains the correct key, which is compared against the input, and a mismatch is rotated into the reject branch.} Write the set of realized keys in binary and read them starting from the most significant bit.  At the first bit where some members of the key set disagree, split the set into subsets and write the bit into an ancilla with a CX. For each subset, proceed to the next bit where some members disagree, and split again at that point. Continue walking the binary tree in this way until every group holds a single key. In the circuit, each leg of the tree is mapped to a clean flag ancilla and the node from which the leg emerges corresponds to a tested bit. A flag ancilla on a ``bit$=$1" leg is computed by a Toffoli on the parent leg's flag ancilla and the node bit. A flag ancilla on a ``bit$=$0" leg is computed by the XOR of the parent leg's flag ancilla and the flag ancilla on the bit$=$1 leg (A AND NOT B = A XOR (A AND B)).  At the top of the tree, the first node, the Toffoli becomes a CX, and the two CX of the XOR become one CX.

This sparse unary iteration flips on exactly one flag for each known key. Conditioned on each flag, the data is written one bit at a time by CX gates.

A tree that fully separates a set of $M$  keys has $M-1$  nodes. The circuit therefore requires $M-2$ Toffolis, or $4(M-2)\,T$. The nice feature is this cost is independent of the width of the keys. This  method to implement a QROM is the counterpart of the  dense unary iteration of~\cite{babbush2018qrom}, which walks the contiguous range of bitstrings $0,\dots,M-1$.  Also, ancillas are heavily reused. The tree is walked depth-first and the flags are uncomputed on retreat, so no more than the key width in bits ($\calO(10)$) independent work qubits are needed.  

The rule that a key already singled out by the tree does not have its remaining bits tested is morally the same optimization as the control pruning of multicontrolled gates used in the earlier circuit constructions for SU(3) gauge theory~\cite{balaji2025circuits}. In that approach, transitions are realized by Givens rotations, and controls are dropped if they are not needed to distinguish the physical bitstrings.  Here that pruning acts at the level of the lookup rather
than of a single gate. 

Another route to implementing a sparse table lookup is to  sort the realized keys classically into a dense counting register, iterate densely over that register, and return the raw key as part of the lookup data~\cite{berry2019qubitization}.   We use the QROM with lookup described above in PREP, but it is simple rather than optimal.

\section{Electric block encoding}
\label{app:electric-be}

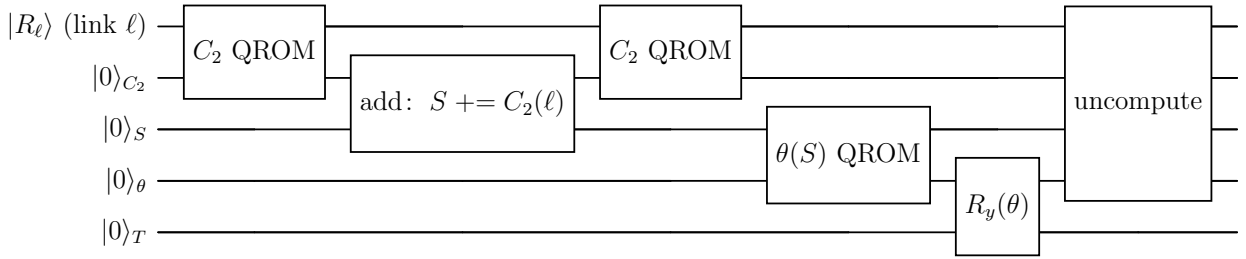
\begin{figure}[htbp]
\centering
\resizebox{\textwidth}{!}{%
\begin{quantikz}[column sep=0.4cm, row sep={0.8cm,between origins}]
\lstick{$|R_\ell\rangle$ (link $\ell$)}
  & \gate[2]{C_2\ \mathrm{QROM}} & \qw & \gate[2]{C_2\ \mathrm{QROM}} & \qw & \qw & \gate[4]{\mathrm{uncompute}} & \qw \\
\lstick{$|0\rangle_{C_2}$}
  & & \gate[2]{\mathrm{add}\!:\ S \mathrel{+}= C_2(\ell)} & & \qw & \qw & & \qw \\
\lstick{$|0\rangle_{S}$}
  & \qw & & \qw & \gate[2]{\theta(S)\ \mathrm{QROM}} & \qw & & \qw \\
\lstick{$|0\rangle_{\theta}$}
  & \qw & \qw & \qw & & \gate[2]{ R_y(\theta)} & & \qw \\
\lstick{$|0\rangle_{T}$}
  & \qw & \qw & \qw & \qw & & \qw & \qw
\end{quantikz}%
}
\caption{Electric block encoding $U_E$ for one link block (repeated
over all links).}
\label{fig:electric}
\end{figure}

$E^2(\ell)$ is proportional to the quadratic Casimir of the irrep on link $\ell$. Therefore the electric Hamiltonian $H_E = \tfrac{g^2}{2a}\sum_\ell E^2(\ell)$ is diagonal in the irrep basis and we can use a relatively simple block encoding (Fig.~\ref{fig:electric}).  Running over the lattice, we feed each link   into a QROM and write the appropriate Casimir into a register $C_2$, then add $C_2$ into a running total in a register $S$, and finally uncompute  $C_2$. Then the sweep moves to the next link. After the sum is finished, a select-swap QROM~\cite{low2019dirty} keyed on $\sim 10\,N_{\rm links}$ values of $S$ loads the angle $\theta(S)=2\arccos\!\big(\tfrac{g^2}{2}S/\alpha_E\big)$ into the $\theta$ register with precision $b_\theta$ bits, and a programmed $R_y(\theta)$ writes $H_E/\alpha_E$ into the flag $T$ qubit amplitude.   On a link irrep eigenstate $|R\rangle$ with electric energy $E_R=\tfrac{g^2}{2}C_2(R)$, the lookups load $S=C_2(R)$ and the rotation takes the flag to $\tfrac{E_R}{\alpha_E}|0\rangle_T+\sqrt{1-(E_R/\alpha_E)^2}\,|1\rangle_T$.  For a general state $|\psi\rangle=\sum_\psi c_\psi|\psi\rangle$, projecting $T$ onto $|0\rangle$ returns $\sum_\psi c_\psi\tfrac{E_\psi}{\alpha_E}|\psi\rangle=H_E|\psi\rangle/\alpha_E$. (The flag $T$ is thus part of the ancilla register $a$ that is projected onto the upper-left block.) The uncompute box is the QROM adjoint on $\theta$ followed by another sweep over the lattice to return $S$ to $|0\rangle$.

This is a simpler instance of the lookup-plus-rotation also used by the amplitude-weighted prepare circuit described in \S\ref{sec:oracle}. Its building blocks are the lookup and programmed rotations discussed in \S\ref{sec:data-quantum}. Because the electric key is a dense contiguous range of total Casimir values, these lookups  can be performed directly with a select-swap QROM~\cite{low2019dirty}, rather than the sparse iteration used by PREP, and its cost scales with the square root of the number of rows.

Alternatively, the electric term might be removed from the block encoding altogether by working in the interaction picture, where $e^{-iH_E\tau}$ is a product of single-link $R_z$ rotations that are fast-forwarded~\cite{low2018interaction,rajput2022hybridized}. This would drop  $\alpha_E$ from the QSVT degree.

\bibliographystyle{utphys}
\bibliography{refs}

\end{document}